\documentclass[12pt]{article}

\usepackage{amssymb, amsmath}
\usepackage{graphicx}
\usepackage{cite}
\usepackage{hyperref}
\usepackage{comment}
\usepackage{xcolor}
\def\be{\begin{equation}}
\def\ee{\end{equation}}
\def\bea{\begin{eqnarray}}
\def\eea{\end{eqnarray}}
\def\bes{\begin{equation*}}
\def\ees{\end{equation*}}
\def\beas{\begin{eqnarray*}}
\def\eeas{\end{eqnarray*}}
\def\mcF{\mathcal{F}}

\def\mcF{\mathcal{F}}

\def\hdel{D}

\def\td{ \textrm{d}}

\newtheorem{theorem}{Theorem}
\def\nn{\nonumber}

\title{ {\bf Electrovacuum spacetime near an extreme horizon}}

\author{Carmen Li$^{a}$\footnote{k.k.li@lancaster.ac.uk} \  and James Lucietti$^{b}$\footnote{j.lucietti@ed.ac.uk } 
\\ \\  \small \sl  $^a$Faculty of Physics, University of Warsaw, ul. Pasteura 5, 02-093 Warsaw, Poland 
\\ \small \sl and School of Computing and Communications, InfoLab21,
 \\ \small \sl Lancaster University, Lancaster, LA1 4WA, UK
\\ \\  \small \sl $^b$  School of Mathematics and Maxwell Institute for Mathematical Sciences, \\ \small \sl  University of Edinburgh, King's Buildings, Edinburgh, EH9 3FD, UK
}

\date{}

\begin{document}

\maketitle

\begin{abstract}
We determine all  infinitesimal transverse deformations of extreme horizons in Einstein-Maxwell theory that preserve axisymmetry. In particular, we show that the general static transverse deformation of the AdS$_2\times S^2$ near-horizon geometry is a two-parameter family, which contains the known extreme charged, accelerating, static black hole solution held in equilibrium by an external electric or magnetic field (Ernst solution) and a special case of the extreme Kerr-Newman-Melvin solution.  More generally, we find a three-parameter family of deformations of the extreme Kerr-Newman horizon, which contains the extreme Kerr-Newman-Melvin solution and a rotating generalisation of the Ernst solution.   We also consider vacuum gravity with a cosmological constant and prove uniqueness of axisymmetric transverse deformations of the extreme Kerr-AdS horizon. Finally, we completely classify transverse deformations of extreme horizons in three-dimensional Einstein-Maxwell theory with a negative cosmological constant.
\end{abstract}

\newpage

\tableofcontents


\section{Introduction and main results}
Extreme horizons in General Relativity in four and higher dimensions possess a number of remarkable rigidity properties. Most notably, it has been established that for stationary and axially symmetric spacetimes containing an extreme (Killing) horizon, the near-horizon geometry possesses an enhanced isometry group $SO(2,1)$, in a wide range of Einstein-Maxwell-scalar type theories  (which include various $D=4,5$ supergravity theories) and Einstein-Yang-Mills theory~\cite{Kunduri:2007vf, Figueras:2008qh, Lucietti:2012sa, Li:2013gca}.   
Furthermore,  a number of explicit classifications of near-horizon geometries have been derived assuming axial or supersymmetry (mostly in $D=4,5$), see~\cite{Kunduri:2013ana} for a review. In addition, general restrictions on the topology of extreme horizons have been established~\cite{Lucietti:2012sa, Khuri:2018rzo}. These are all steps towards the general black hole classification program, which is a major open problem in higher dimensional General Relativity. 

Near-horizon classifications are only possible due to the following special property of extreme horizons. The Einstein equations restricted to an extreme horizon reduce to a set of geometric equations purely for data intrinsic to the horizon; the extrinsic data on the horizon decouples if and only if the horizon is extreme. Indeed, this is why one can define a precise notion of the near-horizon geometry for a spacetime containing an extreme horizon. The Einstein equations thus reduce to a problem of Riemannian geometry on a spatial section of the horizon which is blind to the exterior spacetime. This great simplification is responsible for the above rigidity and classification results. However, it also highlights a major difficulty: given a near-horizon geometry how do you reconstruct the possible corresponding extreme black hole solutions (should they exist)?

Nevertheless, the rigidity of near-horizon geometries provides the key boundary conditions required to establish uniqueness and classification theorems for extreme black holes. This has been accomplished for $D=4$ Einstein-Maxwell black holes~\cite{Chrusciel:2006pc, Amsel:2009et, Figueras:2009ci, Chrusciel:2010gq}, $D=5$ stationary and biaxisymmetric vacuum black holes~\cite{Figueras:2009ci}, supersymmetric black holes to $D=4, N=2$ supergravity~\cite{Chrusciel:2005ve}, and most recently for supersymmetric and biaxisymmetric black holes in $D=5$ minimal supergravity~\cite{Breunholder:2017ubu}. These works all combine global constraints on the spacetime arising from asymptotic flatness and the assumed special symmetry (axial or supersymmetry), together with the near-horizon classification.

Despite these advances, it is desirable to develop a `quasi-local' approach which does not rely on global spacetime assumptions or special symmetry structures.
It is well known that a given near-horizon geometry may not arise as the near-horizon limit of any black hole, or may arise as the near-horizon limit of distinct black hole solutions.
It is therefore of interest to study the more general inverse problem: what are the possible extreme black holes with a given near-horizon geometry?  

In a previous paper we initiated a systematic study of this question in vacuum gravity by introducing the notion of infinitesimal {\it  transverse deformations} of an extreme horizon~\cite{Li:2015wsa}.  These deformations are solutions to the linearised Einstein equations in the background near-horizon geometry (itself a solution due to extremality). This revealed that the moduli space of such deformations, for horizons with compact cross-sections, is finite dimensional -- there cannot be `too many' solutions. Furthermore, by classifying axially symmetric solutions, we established uniqueness of transverse deformations of the extreme Kerr horizon, thereby extending the known rigidity result for the near-horizon geometry~\cite{Lewandowski:2002ua, Kunduri:2008tk}, without invoking any global assumptions on the spacetime.

The purpose of this paper is to extend our study of transverse deformations of extreme horizons to Einstein-Maxwell theories. In fact, following the vacuum theory~\cite{Li:2015wsa}, it has been already shown that the moduli space of transverse deformations to extreme horizons with compact cross-sections is finite dimensional in a large class of Einstein-Maxwell-scalar theories~\cite{ Fontanella:2016lzo, Dunajski:2016rtx}. In this paper we will go further and explicitly determine all axisymmetric deformations in $D=4$ Einstein-Maxwell theory. The results (and analysis) are more complicated than in the vacuum theory.  First, we recall that near-horizon geometries in this theory are known to be unique: the general static solution is AdS$_2\times S^2$ and the general axisymmetric solution is given by the extreme Kerr-Newman horizon~\cite{Lewandowski:2002ua, Kunduri:2008tk}. Our main results are contained in the following theorems:
\begin{theorem}\label{thm:static}
The moduli space of smooth, static and axisymmetric, transverse deformations of the AdS$_2\times S^2$ near-horizon geometry in Einstein-Maxwell theory is  2-dimensional. It contains the extreme Reissner-Nordstr\"om solution, the extreme Ernst solutions and a special case of the extreme Kerr-Newman-Melvin solution.
\end{theorem}
\begin{theorem}\label{thm:KN}
The moduli space of smooth, axisymmetric, transverse deformations of the extreme Kerr-Newman horizon in Einstein-Maxwell theory is  3-dimensional. It contains the extreme Kerr-Newman solution, the extreme Kerr-Newman-Melvin solution and the extreme rotating Ernst solution.
\end{theorem}

\noindent {\it Remarks.} 
\begin{enumerate}
\item The extreme Kerr-Newman-Melvin solution is a 3-parameter family of regular stationary and axisymmetric solutions to Einstein-Maxwell theory, constructed by immersing a Kerr-Newman solution in an external magnetic field, see e.g.\cite{Gibbons:2013yq, Booth:2015nwa, Hejda:2015gna, Bicak:2015lxa}. It occupies a
2-dimensional region of the 3-dimensional moduli space of transverse deformations. It contains a special case in which the near-horizon geometry and deformation are static.
\item The extreme Ernst solution is a 2-parameter family of regular static and axisymmetric solutions to Einstein-Maxwell theory, given by balancing a charged C-metric in an external electric or magnetic field, see e.g.\cite{Griffiths:2009dfa}. In the case of vanishing acceleration parameter it reduces to the Reissner-Nordstr\"om solution. The Ernst solution has a near-horizon geometry isometric to AdS$_2\times S^2$~\cite{Dowker:1994up} (as it must by the above mentioned uniqueness theorem).  We find that although the dimensionality matches, this solution does not fill all parts of the moduli space of static deformations. 
\item A rotating generalisation of the Ernst solution can be constructed from a charged rotating C-metric held in equilibrium by an external magnetic field~\cite{Astorino:2016xiy}. It has a near-horizon geometry isometric to that of the Kerr-Newman black hole~\cite{Astorino:2016xiy} (again this is guaranteed by the near-horizon uniqueness theorem). It occupies a 3-dimensional region of the
3-dimensional moduli space.
\item We also found a 1-parameter family of axisymmetric deformations of AdS$_2\times S^2$ that do not preserve staticity. 
\end{enumerate}

The above results show that the deformations are more general than those corresponding to the Reissner-Nordstr\"om and Kerr-Newman solutions. Hence, we find that the `local' no-hair theorem established in vacuum gravity~\cite{Li:2015wsa} does not persist in Einstein-Maxwell theory, although the physical interpretation of the extra parameters is essentially clear: background electromagnetic field and acceleration.    Indeed, since we do not impose any global assumptions such as asymptotic flatness, regular solutions corresponding to black holes in external electromagnetic fields are naturally captured in our classification. Having said this, as noted in the remarks above, not all regions of the moduli space are occupied by the known solutions, so it would be interesting to determine if these extend to other solutions or if there is some obstruction to extending these linearised solutions to higher order.

We also consider the case of a pure cosmological constant. Uniqueness of axisymmetric near-horizon geometries has been previously established~\cite{Kunduri:2008tk}. Our main result is:
\begin{theorem}\label{thm:ads}
Any smooth, axisymmetric, marginally trapped, transverse deformation of the extreme Kerr-AdS horizon corresponds to an extreme Kerr-AdS black hole.
\end{theorem}

This is interesting because black hole uniqueness theorems for solutions with a cosmological constant have not been established, and indeed may be violated. The violations of uniqueness are expected for solutions with a single Killing field (a combination of the stationary and axial symmetries)~\cite{Kunduri:2006qa, Dias:2015rxy}, so our result is perhaps not surprising. To this end, it would be interesting to extend our analysis to deformations which do not preserve the axial symmetry. On the other hand, for axisymmetric solutions it is possible that the no-hair theorem for Kerr-AdS is indeed true; our results are consistent with this and may be taken as weak evidence for this possibility.   

It is worth recalling that gravitational perturbations of the near-horizon geometry of the extreme Kerr(-AdS) black hole and of AdS$_2\times S^2$ have been previously considered~\cite{Amsel:2009ev, Dias:2009ex, Dias:2012pp, Porfyriadis:2018yag}.  These works consider dynamical perturbations. In contrast, our transverse deformations are perturbations which are non--dynamical since by construction they preserve the structure of the extreme Killing horizon.

Naturally,  it would be interesting to extend our results to Einstein-Maxwell solutions with a cosmological constant. Preliminary work on this suggests the analysis and results are (even) more complicated in this case. Due to the lack of solution generating techniques, black hole solutions in external electromagnetic fields with a cosmological constant are not known. Nevertheless, the classification of near-horizon geometries of extreme horizons has been also solved in this case~\cite{Kunduri:2008tk}. Hence investigating their transverse deformations may provide some evidence for the existence (or nonexistence) of putative dS/AdS black hole solutions in external electromagnetic fields.

Finally, as a simpler application of our general formalism, we consider three dimensional Einstein-Maxwell theory with a negative cosmological constant. This theory admits charged generalisations of the (extreme) BTZ black hole~\cite{Banados:1992wn, Clement:1992ke, Clement:1995zt, Martinez:1999qi}.  In this case we show that one can easily determine all transverse deformations of the possible extreme horizons in the theory.

This paper is organised as follows.  In Section \ref{sec:EHEM} we derive the linearised Einstein-Maxwell equations for a general transverse deformation of an extreme horizon in $D$-dimensional spacetime. In Section \ref{sec:modulispaces} we determine the general axisymmetric deformations of AdS$_2 \times S^2$ and more generally of the extreme Kerr-Newman horizon. In Section \ref{sec:ads} we prove uniqueness of deformations of static extreme horizons with a cosmological constant and we prove uniqueness of axisymmetric deformations of the extreme Kerr-AdS horizon. In Section \ref{sec:3d} we determine all deformations of extreme horizons in three-dimensional Einstein-Maxwell theory with a negative cosmological constant. In the Appendix we give a general method for computing the transverse deformation arising from an axisymmetric extreme black hole solution and apply it to the known examples.

\section{Near an extreme horizon in Einstein-Maxwell theory}
\label{sec:EHEM}

Let $(M,g)$  be a $D$-dimensional spacetime containing a smooth \textit{degenerate} Killing horizon $\mathcal{H}$ of a complete Killing field $n$ possessing a cross-section $S$ (a $(D-2)$-dimensional spacelike submanifold transverse to $n$). In the neighbourhood of such a horizon we introduce Gaussian null coordinates (GNC) and the associated near-horizon geometry, see e.g.~\cite{Kunduri:2013ana} for more details. The spacetime metric in these coordinates takes the form
\be
g = 2  \td v \left(  \td r + r h_a (r, x)  \td x^a +\tfrac{1}{2} r^2 F(r, x) \td v \right) + \gamma_{ab} (r, x) \td x^a \td x^b \; ,  \label{gnc}
\ee
where the horizon is at $r=0$, the vector field $\ell= \partial_r$ is transverse and geodesic, the normal Killing field $n= \partial_v$, and $(x^a)$ are coordinates on a cross-section $S$. Degeneracy of the horizon is equivalent to $g_{vv}= O(r^2)$.  From this form of the metric, it is easy to see there exists a well defined notion of a near-horizon geometry.

For any $\varepsilon >0$, consider the diffeomorphism $\phi_\varepsilon : (v,r,x^a) \to ( v / \varepsilon,  \varepsilon r , x^a)$ and define the 1-parameter family of metrics 
\be
g(\varepsilon)\equiv \phi^*_{\varepsilon} g = 2  \td v \left(   \td r + r h_a (\varepsilon r, x) \td x^a +\tfrac{1}{2} r^2  F( \varepsilon r, x) \td v\right) + \gamma_{ab} (\varepsilon r, x) \td x^a \td x^b \;  .     \label{spacetime}
\ee
The near-horizon geometry is defined as the $\varepsilon \rightarrow 0$ limit of $g(\varepsilon)$, which we denote by $g^{(0)}$. Smoothness of the metric guarantees it exists  and it is given by 
\be
g^{(0)}= 2 \td v \left( \td r + r h^{(0)}_a (x) \td x^a +\tfrac{1}{2 } r^2  F^{(0)}(x) \td v\right) + \gamma^{(0)}_{ab} (x) \td x^a \td x^b  \; , \label{nhg}
\ee
where $F^{(0)}= F|_{r=0}$, $h^{(0)}_a(x)= h_a|_{r=0}$, $\gamma^{(0)}_{ab}(x) = \gamma_{ab}|_{r=0}$, are a function, 1-form and Riemannian metric on $S$. In particular, $\gamma^{(0)}_{ab}(x)$ is the induced metric on $S$. 

The transverse deformation of an extreme horizon, introduced in~\cite{Li:2015wsa}, is defined as the first variation of $g(\varepsilon)$ at $\varepsilon=0$, i.e. $g^{(1)}\equiv  \frac{d}{d\varepsilon} g(\varepsilon)|_{\varepsilon=0}$. This is equivalent to a linear perturbation of the near-horizon geometry and is explicitly given by
\be
g^{(1)} = r^3 F^{(1)}(x) \td v^2 + 2r^2 h^{(1)}_{a}(x) \td v \td x^a + r \gamma^{(1)}_{ab}(x) \td x^a \td x^b \; ,
\ee
where  $\gamma^{(1)}_{ab} = \partial_r \gamma_{ab}|_{r=0}$ etc. 
The quantity $\gamma^{(1)}_{ab}(x)$ is (twice) the extrinsic curvature of $S$ with respect to the null normal $\ell$.  As for any linearised perturbation, diffeomorphism invariance implies the gauge freedom $g^{(1)}\to g^{(1)} - \mathcal{L}_\xi g^{(0)}$, where $\xi$ is a vector field.  The most general such $\xi$ which preserves the form of the deformation (and is not an isometry of $g^{(0)}$) is~\cite{Li:2015wsa}
\be
\xi  =  \tfrac{1}{2} f \partial_v +\tfrac{1}{4} r^2 h^{(0)}_a D^a f \; \partial_r - \tfrac{1}{2} r D^a f \;  \partial_a  \; ,
 \label{xi}
\ee
where $D_a$ is the metric connection of $\gamma_{ab}^{(0)}$ and $f(x)$ is a smooth function on $S$. Thus $f$ parameterises a supertranslation on the null surface $\mathcal{H}$. In terms of this the gauge transformation rules for the first order data are
\begin{eqnarray}
&&{\gamma}^{(1)}_{ab} \to\gamma^{(1)}_{ab} + D_a D_b f - h^{(0)}_{(a} D_{b)} f  \label{gautransgamma}\\
\nn &&{h}^{(1)}_a  \to h^{(1)}_a  - \tfrac{1}{2} F^{(0)} D_a f - \tfrac{1}{4} (D_a h^{(0)}_{b})  D^b f - \tfrac{1}{4} h^{(0)}_a h^{(0)}_b D^b f + \tfrac{1}{2} (D_b h^{(0)}_a )D^b f + \tfrac{1}{4} h^{(0)}_b D_a D^b f \label{gautransh} \\
&&{F}^{(1)}  \to  F^{(1)} + \tfrac{1}{2} (D^a f) \left( D_a F^{(0)} - h^{(0)}_a F^{(0)} \right) \label{gautransf}   \; . \nn  
\end{eqnarray}
We emphasise that the Gaussian null chart is fixed by a choice of cross-section $S$ and coordinates on $S$. The above gauge freedom emerges as we are effectively requiring the metric to be of Gaussian null form only to first order in the affine parameter $r$.

Now suppose $(M,g)$ is solution to the $D$-dimensional Einstein-Maxwell-$\Lambda$ equations
\bea
&&R_{\mu\nu} =  2 \mcF_{\mu \sigma} \mcF_{\nu}^{~\sigma} - \frac{1}{(D-2)} g_{\mu \nu} \mcF_{\rho \sigma} \mcF^{\rho \sigma} + \Lambda g_{\mu\nu} \\
&&\td \ast \mcF = 0 \; , \qquad \td \mcF=0 \; 
\eea
where $\mathcal{F}$ is the Maxwell 2-form.  In Gaussian null coordinates, a smooth Maxwell field takes the general form\footnote{In general the $\mcF_{va}$ term does \textit{not} admit a near horizon limit; however since on a Killing horizon $0=R_{\mu \nu} K^\mu K^\nu|_\mathcal{N} = 2 \gamma^{ab} \mcF_{va} \mcF_{vb}|_{r=0}$, smoothness implies we must have $ \mcF_{va} = r W_a$ for some smooth $W_a (r,x)$. }
\be
\mcF = \Psi(r,x) \td v \wedge \td r  + r W_a(r,x) \td v \wedge \td x^a + Z_a(r,x) \td r \wedge \td x^a + \tfrac{1}{2} B_{ab} (r,x) \td x^a \wedge \td x^b  \; .
\ee
Define the 1-parameter family of Maxwell fields  $ \mcF (\varepsilon) \equiv \phi_{\varepsilon}^* \mcF$ where $\phi_\varepsilon$ is the diffeomorphism defined above. The near-horizon limit of the Maxwell field is the $\varepsilon \to 0$ limit of  $\mcF (\varepsilon)$. By smoothness of $\mcF$ its near-horizon limit exists and is given by
\be
\mcF^{(0)} = \Psi^{(0)}(x) \td v \wedge \td r  + r W^{(0)}_a(x) \td v \wedge \td x^a + \tfrac{1}{2} B^{(0)}_{ab}(x) \td x^a \wedge \td x^b  \; ,
\ee
where $\Psi^{(0)}= \Psi|_{r=0}$, $W^{(0)}_a= W_a|_{r=0}$, $B^{(0)}_{ab}= B_{ab}|_{r=0}$ are a function, 1-form and 2-form on $S$.  The near-horizon Bianchi identity reduces to 
\be
W^{(0)}_a = D_a \Psi^{(0)}, \qquad  \td B^{(0)}=0.
\ee
The 2-form $B^{(0)}_{ab}$ is the Maxwell field induced on $S$.

The 1-parameter family $(g(\varepsilon), \mcF(\varepsilon))$ must also satisfy the Einstein-Maxwell equations. In particular, the near-horizon limit $(g^{(0)}, \mcF^{(0)})$ satisfies the Einstein-Maxwell equations, resulting in geometric equations for the horizon data $(F^{(0)},h^{(0)}_a, \gamma^{(0)}_{ab}, \Psi^{(0)}, B^{(0)}_{ab})$ intrinsic to $S$. The near-horizon Einstein equations are
\bea
R^{(0)}_{ab} &=& \tfrac{1}{2} h^{(0)}_a h^{(0)}_b - D_{(a} h^{(0)}_{b)}+\Lambda \gamma^{(0)}_{ab} + 2 B^{(0)}_{ac} B^{(0)c}_b + \tfrac{2}{D-2} \gamma^{(0)}_{ab} \Psi^{(0)2} - \tfrac{1}{D-2} \gamma^{(0)}_{ab} B^{(0) 2}  \label{NHeqab}\\
F^{(0)} &=& \tfrac{1}{2} h^{(0)2} - \tfrac{1}{2}D^a h^{(0)}_a+ \Lambda - 2 \left(\tfrac{D-3}{D-2} \right) \Psi^{(0)2} - \tfrac{1}{D-2} B^{(0) 2}
\label{NHeq+-}
\eea
and the near-horizon Maxwell equation is
\be 
\left( \hdel_a - h^{(0)}_a \right) \Psi^{(0)} + \left( \hdel^b - h^{(0)b} \right) B^{(0)}_{ba} =0  \; .
\ee
The classification of solutions to these equations has been extensively studied~\cite{Kunduri:2013ana}. In particular, for $D=4$ all static solutions and all axially symmetric solutions can be determined~\cite{Lewandowski:2002ua, Chrusciel:2005pa, Chrusciel:2006pc, Kunduri:2008rs, Kunduri:2008tk, Li:2013gca}.

Now, we define the transverse deformation of the Maxwell field as the first variation of the Maxwell field $\mcF^{(1)} \equiv \frac{\td \mcF(\varepsilon)}{\td \varepsilon} |_{\varepsilon=0} $. In GNC this reads
\be
\mcF^{(1)}  =  r \Psi^{(1)}(x) \td v \wedge \td r  + r^2 W^{(1)}_a(x) \td v \wedge \td x^a +  Z^{(1)}_a(x) \td r \wedge \td x^a + \tfrac{1}{2} r B^{(1)}_{ab} (x) \td x^a \wedge \td x^b  \; . 
\ee
where $\Psi^{(1)}= \partial_r \Psi|_{r=0}$ etc are defined as above, except $Z^{(1)}_a = Z_a|_{r=0}$ instead.
The linearised Bianchi identity  $\td \mcF^{(1)} =0$
reduces to
\be 
W^{(1)} = \tfrac{1}{2} {\td} \Psi^{(1)} \; , \qquad B^{(1)} = {\td} Z^{(1)} \; .
\ee
The diffeomorphism invariance implies the linearised Maxwell field transforms simultaneously with the linearised metric as $\mcF^{(1)}\to  \mcF^{(1)} - \mathcal{L}_\xi \mcF^{(0)}$, where $\xi$ is the vector field (\ref{xi}).
This implies the first order data transforms as 
\bea
&& \Psi^{(1)} \to \Psi^{(1)} + \left(D_a \Psi^{(0)}- 2 h^{(0)}_a \Psi^{(0)} \right) D^a f  \nn\\
&& Z^{(1)}_a\to{Z}^{(1)}_a + \tfrac{1}{2} \left(\Psi^{(0)} \hdel_a f - B^{(0)}_{ab} \hdel^b f  \right)  \; . \label{gautransZ}
\eea
The linearised Maxwell equation is $\frac{d}{d\varepsilon} ( g^{\mu\nu}(\varepsilon) \nabla_{\mu}(\varepsilon) \mcF_{\nu \rho}(\varepsilon))|_{\varepsilon=0}=0$. 

After some lengthy calculations we find that the linearised Einstein-Maxwell equations reduce to a pair of linear PDEs for first order data $(\gamma^{(1)}_{ab}, Z_a^{(1)})$, defined on the background horizon $(S, \gamma^{(0)}_{ab}, h^{(0)}_a, \Psi^{(0)}, B^{(0)}_{ab})$,  given by
\bea 
\nn 0&=&  \Delta_L \gamma^{(1)}_{ab} + \tfrac{1}{2} {\hdel}_{(a} {\hdel}_{b)}\gamma^{(1)}  + \tfrac{3}{2} h^{(0)c} {\hdel}_c \gamma^{(1)}_{ab} - \tfrac{3}{2} h^{(0)}_{(a} {\hdel}_{b)}\gamma^{(1)}  - h^{(0)c} {\hdel}_{(a}  \gamma^{(1)}_{b)c} + h^{(0)}_{(a} {\hdel}^c  \gamma^{(1)}_{b)c} -  \tfrac{1}{2} h^{(0)2} \gamma^{(1)}_{ab}  \\
\nn  && + \tfrac{1}{2} h^{(0)}_a h^{(0)}_b \gamma^{(1)}  + \left( {\hdel}_{(a} {h^{(0)c}}\right) \gamma^{(1)}_{b)c} - \left( \hdel^c h^{(0)}_{(a}\right) \gamma^{(1)}_{b)c} 
+ 2  B^{(0)}_{ac} B^{(0)}_{bd} \gamma^{(1)cd}- \tfrac{2}{D-2} \gamma^{(0)}_{ab} B^{(0)e}_{~c} B^{(0)}_{de} \gamma^{(1)cd} \\
\nn && - 2 \Psi^{(0)2} \gamma^{(1)}_{ab} + \tfrac{2}{D-2} \gamma^{(0)}_{ab} \Psi^{(0)2} \gamma^{(1)} + 4 \Psi^{(0)} \hdel_{(a} Z^{(1)}_{b)} - \tfrac{4}{D-2} \gamma^{(0)}_{ab} \Psi^{(0)} \hdel \cdot Z^{(1)} + 4 \hdel_c Z^{(1)}_{(a} B^{(0)c}_{~b)} \\
\nn & & + \tfrac{4}{D-2} \gamma^{(0)}_{ab} \hdel_{[c} Z^{(1)}_{d]} B^{(0)cd} - 4 \Psi^{(0)} h^{(0)}_{(a} Z^{(1)}_{b)} - 4 Z^{(1)c} h^{(0)}_{(a} B^{(0)}_{b)c} + 4 Z^{(1)c} \hdel_{(a} B^{(0)}_{b)c} -4 Z^{(1)}_{(a} B^{(0)}_{b)c} h^{(0)c} \\
&& + \tfrac{4}{D-2} \gamma^{(0)}_{ab} Z^{(1)}_c \hdel^c \Psi^{(0)} + \tfrac{4}{D-2} \gamma^{(0)}_{ab} B^{(0)cd} Z^{(1)}_{[c} h^{(0)}_{d]} \label{einstein}  \; ,
\eea
 where  $\Delta_L \gamma^{(1)}_{ab} =- \tfrac{1}{2} D^2  \gamma^{(1)}_{ab} + {R}_{(a}^{(0) c} \gamma^{(1)}_{b)c} - {R}_{acbd}^{(0)}\gamma^{(1)cd}$ is the Lichnerowicz operator of $(S, \gamma^{(0)}_{ab})$, and
\bea
\nn 0 &=& \hdel^2 Z^{(1)}_a - \hat{R}^{(0)}_{ab} Z^{(1)b} + h^{(0)b} \hdel_a Z^{(1)}_b - 3 h^{(0)b} \hdel_b Z^{(1)}_a - h^{(0)}_a \hdel \cdot Z^{(1)} + 4 B^{(0)}_{ab} B^{(0)bc} Z^{(1)}_c \\
\nn & &  - 4 \Psi^{(0)2} Z^{(1)}_a - 2 F^{(0)} Z^{(1)}_a + 2 h^{(0)2} Z^{(1)}_a  + 2 Z^{(1)b} \hdel_{[b} h^{(0)}_{a]} - Z^{(1)}_a \hdel \cdot h^{(0)} - B^{(0)}_{ab} \hdel_c \gamma^{(1)bc} \\
\nn & & + \Psi^{(0)} \hdel^{b} \gamma^{(1)}_{ab} - \gamma^{(1)bc} \hdel_b B^{(0)}_{ca} - \gamma^{(1)bc} B^{(0)}_{ab} h^{(0)}_c - \gamma^{(1)}_{ab} W^{(0)b} - \gamma^{(1)}_{ab} B^{(0)b}_{~~~~c} h^{(0)c}  \\
& & - \gamma^{(0)bc} \hat{C}^{(1)d}_{bc} B^{(0)}_{da} - \gamma^{(0)bc} \hat{C}^{(1)d}_{ab} B^{(0)}_{cd} + B^{(0)}_{ab} \hdel^b \gamma^{(1)} - \tfrac{3}{2} \Psi^{(0)} \hdel_a \gamma^{(1)} + h^{(0)}_a \Psi^{(0)} \gamma^{(1)} \label{maxwell} \; ,
\eea
where $ \hat{C}^{(1)c}_{ab} = \tfrac{1}{2} \gamma^{(0) cd} ( D_a \gamma^{(1)}_{db}+ D_b \gamma^{(1)}_{ad}- D_d \gamma^{(1)}_{ab})$. The rest of the first order data $(F^{(1)}, h^{(1)}_a, \Psi^{(1)})$  is then determined algebraically by
\bea
h^{(1)}_a &=& \tfrac{1}{2} h^{(0)b} \gamma^{(1)}_{ab} - \tfrac{1}{2} \hdel^b  \gamma^{(1)}_{ab} + \tfrac{1}{2} \hdel_a \gamma^{(1)} - \tfrac{1}{4} h^{(0)}_a \gamma^{(1)}+ 2 \left(B^{(0)b}_{~a} Z^{(1)}_b + \Psi^{(0)} Z^{(1)}_a  \right) \label{E-a} \\
F^{(1)} &=&  h^{(0)a} h_a^{(1)} - \tfrac{1}{3} {D}^a h_a^{(1)} - \tfrac{1}{3} h^{(0)a} h^{(0)b} \gamma^{(1)}_{ab}  + \tfrac{1}{6} h^{(0) (a} {D}^{b)} \gamma^{(1)}_{ab} \label{E+-}  \\
& &+ \tfrac{1}{6}\left( {D}^{(a} h^{(0)b)} \right) \gamma^{(1)}_{ab}  - \tfrac{1}{6} F^{(0)}  \gamma^{(1)}  - \tfrac{1}{12} h^{(0)a} \left( {D}_a  \gamma^{(1)}   - h^{(0)}_a  \gamma^{(1)}  \right)  \nn \\  &&- \tfrac{4}{3} \left(\tfrac{D-3}{D-2}\right) \Psi^{(0)} \Psi^{(1)} - \tfrac{2}{3} \left( \tfrac{D-4}{D-2}\right) \Psi^{(0)} h^{(0)a} Z^{(1)}_a + \tfrac{2}{3} \left( \tfrac{D-4}{D-2} \right) W^{(0)a} Z^{(1)}_a  \nn \\ 
&&  - \tfrac{2}{3(D-2)} B^{(0)[ab]} \left( B^{(1)}_{[ab]} + 2 Z^{(1)}_{[a} h^{(0)}_{b]} \right) + \tfrac{2}{3(D-2)} \gamma^{(0)cd} B^{(0)}_{[ac]} B^{(0)}_{[bd]} \gamma^{(1)ab} \nn \\ 
 \Psi^{(1)} &=& - h^{(0)a} Z^{(1)}_a + \hdel^a Z^{(1)}_a - \tfrac{1}{2} \gamma^{(1)} \Psi^{(0)} \; .  \label{maxr}
 \eea
 Equations (\ref{einstein}), (\ref{E-a}), (\ref{E+-}) correspond to the Einstein equations, whereas (\ref{maxwell}) and (\ref{maxr}) correspond to the Maxwell equations (although simplified by combining them). 
 Notice that (\ref{einstein}) is automatically traceless, so the number of independent PDEs given by (\ref{einstein}) and (\ref{maxwell}) is the same as the number of degrees of freedom $(\gamma^{(1)}_{ab}, Z^{(1)}_a)$, once the gauge freedom (\ref{gautransgamma}), (\ref{gautransZ}) is accounted for (recall this is parameterised by one function $f$). 
Crucially, (\ref{einstein}) and (\ref{maxwell}) are linear elliptic (once gauge fixed) PDEs for the first order data $(\gamma^{(1)}_{ab}, Z^{(1)}_a)$ defined on the background of the horizon geometry $(S, \gamma^{(0)}_{ab}, h^{(0)}_a, \Psi^{(0)}, B^{(0)}_{ab})$.   By application of standard results for Fredholm operators we deduce: \\

{\it  \noindent
The moduli space of transverse deformations of a near-horizon geometry of an extremal horizon with compact cross-sections in $D$-dimensional Einstein-Maxwell-$\Lambda$ theory is finite dimensional.}  \\

This result was established for vacuum gravity in~\cite{Li:2015wsa} and subsequently generalised to a large class of Einstein-Maxwell-dilaton theories in~\cite{Fontanella:2016lzo}.

\section{Moduli spaces of electrovacuum deformations}
\label{sec:modulispaces}

In $D=4$ Einstein-Maxwell theory, an essentially complete understanding of the possible near-horizon geometry of extreme horizons with compact cross-sections has been achieved. In this section we will set the cosmological constant $\Lambda=0$. The general static near-horizon geometry is AdS$_2\times S^2$~\cite{Chrusciel:2006pc}, whereas the general axisymmetric near-horizon geometry is that of the extreme Kerr-Newman black hole~\cite{Lewandowski:2002ua, Kunduri:2008tk}. We will determine the complete moduli space of transverse deformations which preserve the axisymmetry of these near-horizon geometries.

\subsection{$AdS_2\times S^2$}

This near-horizon geometry of course arises as the near-horizon limit of the extreme Reissner-Nordstr\"om solution. It can be written as
\bea
g^{(0)} &=& -\frac{r^2}{r_+^2} \td v^2 + 2 \td v \td r + r_+^2 \left( \frac{ \td x^2}{1-x^2}  + (1-x^2) \td \phi^2  \right) \\
\mcF^{(0)} &=& - \frac{q_e}{r_+^2} \td v \wedge \td r +  q_m \td x \wedge \td \phi  \nn  \; ,
\eea
where the electric $q_e$ and magnetic $q_m$ charges satisfy $q_e^2 + q_m^2 = r_+^2>0$.  The metric in the round brackets is just the unit $S^2$ written in polar coordinates, so $-1<x< 1$ and $\phi$ is $2\pi$ periodic and $x =\pm 1$ are the usual coordinate singularities corresponding to the poles of $S^2$. We will consider axisymmetric transverse deformations to this near-horizon geometry, i.e., we assume the first order data is invariant under the axial Killing field $\partial_\phi$.

The gauge freedom for such deformations must be generated by an axisymmetric function $f(x)$~\cite{Li:2015wsa}. The gauge transformations (\ref{gautransgamma}) and (\ref{gautransZ}) are
\bea
\gamma^{(1)}_{xx} & \rightarrow &  \gamma^{(1)}_{xx} - \frac{x}{1-x^2} f'(x) + f''(x) \nn \\
\gamma^{(1)}_{x \phi} & \rightarrow &  \gamma^{(1)}_{x \phi}  \nn \\
\gamma^{(1)}_{\phi \phi} & \rightarrow &  \gamma^{(1)}_{\phi \phi} - x (1-x^2) f'(x) \nn  \\
Z^{(1)}_{x} & \rightarrow &  Z^{(1)}_{x} -\frac{q_e }{2 r_+^2}\; f'(x) \nn \\
Z^{(1)}_{\phi} & \rightarrow &  Z^{(1)}_{\phi} -\frac{q_m \left(x^2-1\right)}{2 r_+^2}\; f'(x) \; . 
\eea
Thus $\gamma^{(1)}_{x \phi}$  is gauge invariant.  Using the above transformations we may also define the following gauge invariant variables
\bea
Q_1 &=& q_e Z^{(1)}_\phi + q_m (1-x^2) Z^{(1)}_x \nn \\
Q_2 &=& x^2 (1-x^2)^2 \gamma^{(1)}_{xx} + x(1-x^2) \gamma^{(1)'}_{\phi \phi} + (2 x^2 -1) \gamma^{(1)}_{\phi \phi} \nn \\
Q_3 &=& 2 q_m x Z^{(1)}_\phi - 2  q_e x (1-x^2) Z^{(1)}_x + \gamma^{(1)}_{\phi  \phi} \label{RNgiquant}  \; .
\eea
It is convenient to express the linearised Einstein-Maxwell equations in terms of these gauge invariant quantities.  It is important to note that $Q_1, Q_2, Q_3$ are all globally defined functions on $S^2$ which vanish at the poles $x=\pm 1$. This is easy to see by writing the $Q_i$ in terms of the globally defined vector fields $m=\partial_\phi$ and $X=(1-x^2)\partial_x$ which both vanish at the poles\footnote{For example $Q_2 = x^2\gamma^{(1)}(X,X)+ x X[ \gamma^{(1)}(m,m)] + (2x^2-1) \gamma^{(1)}(m,m)$.}.  It can also be shown that $\gamma^{(1)}_{x\phi}$ is globally defined and vanishes at the poles~\cite{Li:2015wsa}.

The linearised Maxwell equations (\ref{maxwell}) and Einstein equations (\ref{einstein}) reduce to 4 ODEs (two from the Maxwell equations and two from the Einstein equations, due to the latter being traceless). Writing these in terms of the gauge invariant quantities we find
\bea
0&=& 2 q_m x^3 (1-x^2) \left(  \left[\left(1-x^2\right) \gamma^{(1)}_{x \phi} \right]'-\left(1-x^2\right) Q_1''+2 Q_1  \right) \label{RNME1}\\
\nn & - &q_e \left[ x (1-x^2)^2 \left( - x Q_3'' + 2 Q_3' \right) + 2 (1-x^2)(2x^2 - 1) Q_3 +x (1-x^2) Q_2' + 2 (3x^2-1)Q_2\right]   \\
0 &=& \left(1-x^2\right)  \left[ (1 - x^2 ) \gamma^{(1)}_{x \phi} \right]'' +  2 x \left[  (1-x^2) \gamma^{(1)}_{x \phi}  \right]' + 2 (1-x^2) \gamma^{(1)}_{x \phi}  \label{RNR1XP} \\  
&+& 4 (1-x^2) Q_1' + 8 x Q_1\nn \\
0 &=& 2 q_e x^3(1-x^2) \left( \left[ (1-x^2) \gamma^{(1)}_{x \phi} \right]'    - (1-x^2) Q_1'' + 2 Q_1 \right)   \label{RNME2}  \\
\nn &+&q_m \left[ x (1-x^2)^2 \left( - x Q_3'' + 2 Q_3' \right) + 2 (1-x^2)(2x^2 - 1) Q_3 + x (1-x^2) Q_2' + 2  (3x^2-1)Q_2 \right] \\
 0 &=& -x (1-x^2) Q_2' + 2 (1-3x^2) Q_2 - 2 x (1-x^2)^2 Q_3' + 2 (1-x^2)(1-3x^2) Q_3 \label{RNR1XX} \; . 
\eea
Note that equations (\ref{RNME1}) and (\ref{RNME2}) imply that the coefficients of $q_e$ and $q_m$ in each must vanish separately (since $q_e^2+q_m^2 \neq 0$) and are thus equivalent to
\bea
0 &=& \left[\left(1-x^2\right) \gamma^{(1)}_{x \phi} \right]'-\left(1-x^2\right) Q_1''+2 Q_1 
\label{RNME12} \\
0&=&  x (1-x^2)^2 \left( - x Q_3'' + 2 Q_3' \right) + 2 (1-x^2)(2x^2 - 1) Q_3  \nn \\
 && +x (1-x^2) Q_2' + 2 (3x^2-1)Q_2 \label{RNME21}  \; .
\eea
Thus the pairs of variables $(\gamma_{x\phi}^{(1)}, Q_1)$ and $(Q_2, Q_3)$ decouple, each satisfying a pair of ODEs. 

Let us first consider the equations for $(\gamma_{x\phi}^{(1)}, Q_1)$ which are \eqref{RNME12} and \eqref{RNR1XP}. We can obtain an expression for $\left[\left(1-x^2\right) \gamma^{(1)}_{x \phi} \right]'$ in terms of $Q_1$ and its derivatives from \eqref{RNME12}, and substitute this expression into \eqref{RNR1XP} to solve for $\gamma^{(1)}_{x \phi} $ in terms of $Q_1$ and its derivatives only
\be 
\gamma^{(1)}_{x \phi} = \frac{ (1-x^2)^2 Q_1''' + 2 (1-x^2) Q_1' + 4 x Q_1 }{ -2  \left(1-x^2\right)} \label{gxp1inq1} \; . 
\ee
Substituting this back into the first equation \eqref{RNME12} gives a fourth order equation for $Q_1$
\be 
0 = - \tfrac{1}{2} (1-x^2) Q_1^{(4)} + 2 x Q_1'''- 2 Q_1'' \; , 
\ee
which has the general solution 
\be 
Q_1 = \tfrac{C_1}{6} x^3 + \tfrac{C_2}{8} x \left( - 2 x + (1-x^2) \log[1-x] - (1-x^2) \log[1+x] \right) + C_3 + C_4 x  \; , 
\ee
where the $C_i$'s are constants. Smoothness of $Q_1$ and the endpoints $x=\pm 1$ requires $C_2=0$. Then \eqref{gxp1inq1} gives
\be
\gamma^{(1)}_{x \phi} = \frac{ C_1(2 x^4 - 3 x^2 +3) + 12 C_3 x + 6 C_4 (1+x^2)}{-6(1-x^2)} \; , 
\ee
which is not smooth at the endpoints in general. Imposing smoothness at $x= \pm 1$ then yields the constraints $C_3 =0$ and $C_1 = - 6 C_4$. Thus by relabelling the constant $C_4 \to K_1$, the general solutions for $Q_1$ and $\gamma^{(1)}_{x \phi}$ are  simply
\bea
Q_1 &=& K_1 x( 1-x^2)  \label{Q1static} \\
\gamma^{(1)}_{x \phi} &=& 2 K_1 (1-x^2) \label{gxpstatic}   \; . 
\eea

Now let us consider the equations for $(Q_2, Q_3)$ which are \eqref{RNME21} and \eqref{RNR1XX}. It is easy to see that the $Q_2$ terms can be eliminated by adding \eqref{RNR1XX} to \eqref{RNME21}, resulting in  
\be 
0 = (1-x^2) Q_3'' + 2 Q_3 \; , 
\ee 
whose general solution which is smooth at $x=\pm 1$ is
\be 
Q_3 = K_2(1-x^2) \;,   \label{Q3static}
\ee
where $K_2$ is a constant.  
Substituting this into \eqref{RNR1XX} gives the following equation for $Q_2$: 
\be 
0 = 2 (1-3x^2) Q_2 - x(1-x^2) Q_2' + 2 K_2 (1-x^2)^3 \; ,  
\ee
which has the general solution
\be 
Q_2 = - K_2 (1-x^2)^2 + K_3 x^2 (1-x^2)^2   \label{Q2static}
\ee
where $K_3$ is another constant. 

To summarise, the general  smooth solution for $\gamma^{(1)}_{x\phi}, Q_1, Q_2, Q_3$ is given by (\ref{gxpstatic}, \ref{Q1static}, \ref{Q2static}, \ref{Q3static}).
Using \eqref{RNgiquant}, we may now invert this general solution for the gauge invariant variables,  to obtain the general solution for individual metric components $\gamma^{(1)}_{ab}$ and Maxwell field components $Z^{(1)}_a$. Because of the gauge freedom, one of them will be a free smooth function. Regularity at the poles implies that without loss of generality, we can write 
\be  
\gamma^{(1)}_{\phi \phi} = (1-x^2) \left( m + x g(x) \right) \; ,
\ee 
where $m$ is a constant and $g(x)$ is some smooth function. Then the gauge transformation for $\gamma^{(1)}_{\phi \phi}$ is simply $g(x) \rightarrow g(x) - f'(x)$. We then find
\bea 
\nn \gamma^{(1)}_{xx} &=& \frac{m-K_2  + x^2(K_2 + K_3) - x^2(1-x^2) g'(x) + x^3 g(x) - K_3 x^4}{x^2(1-x^2)} \\
\nn Z^{(1)}_x &=& \frac{q_e(m - K_2) + 2 K_1 q_m x^2 + q_e x g(x)}{2 (q_e^2 + q_m^2) x} \\
Z^{(1)}_\phi &=& (1-x^2) \frac{q_m(K_2-m ) + 2 K_1 q_e x^2 - q_m x g(x)}{2 (q_e^2 + q_m^2) x} \label{RNsoln} \; . 
\eea   
The solution \eqref{RNsoln} must be smooth for all $-1<x<1$. Therefore, to avoid the pole at $x=0$ we must set $K_2=m$. Then, the general solution is 
\bea 
\nn \gamma^{(1)}_{xx} &=&  \frac{ K_2 + x g(x)}{(1-x^2)}+K_3-g'(x) \\ 
\nn \gamma^{(1)}_{x \phi} &=& 2 K_1 (1-x^2) \\
\nn \nn \gamma^{(1)}_{\phi \phi} &=& (1-x^2)\left( K_2 + x g(x) \right)  \\
\nn Z^{(1)}_x &=& \frac{ 2 K_1 q_m x + q_e  g(x)}{2 (q_e^2 + q_m^2)} \\
Z^{(1)}_\phi &=& (1-x^2) \frac{ 2 K_1 q_e x - q_m  g(x)}{2 (q_e^2 + q_m^2)} \label{RNsoln1} \; . 
\eea   
Observe that near the poles $x^2 \rightarrow 1$ we have 
\be
\gamma^{(1)}_{xx} = \gamma^{(1)}_{\phi \phi} (1-x^2)^{-2} + \mathcal{O}(1) \; , \qquad \gamma^{(1)}_{x\phi}=O(1-x^2)  \; , \qquad \gamma^{(1)}_{\phi\phi}=O(1-x^2) \; ,
\ee
 so this family of deformations $\gamma^{(1)}_{ab}$ is indeed smooth tensor field on $S^2$~\cite{Li:2015wsa}.  In summary, we have found a three parameter $K_1, K_2, K_3$ family of smooth axisymmetric transverse deformations of AdS$_2\times S^2$. Since these linear deformations are defined only up to an overall scale, there are only two independent physical deformations.

Let us now consider the conditions for the deformation to correspond to a marginally trapped surface (MTS); see~\cite{Li:2015wsa} for a full discussion of extreme MTS. The mean transverse expansion, which is a gauge invariant, is given by
\bea
\int_S \gamma^{(1)}= 2 \pi r_+^2 \int^1_{-1} \gamma^{(1)} \; \td x &=& 
		2\pi \int^1_{-1}  2K_2 + K_3 (1-x^2) + \left( (x^2 -1) g(x)\right)' \; \td x  \nn \\
&= & 8\pi \left(K_2 + \frac{K_3}{3} \right)			 \; , 	
\eea
so the MTS condition $\int_S \gamma^{(1)}>0$ is satisfied if 
\be 
K_2 + \frac{K_3}{3} > 0 \; .   \label{MTSstatic}
\ee
 For the extreme Reissner-Nordstr\"om black hole solution it is easy to check that 
 \be
 K_1=K_3=0, \qquad K_2 = 2 r_+.  
\ee
Thus, any solution with $K_1=K_3=0$ which obeys the MTS condition is equivalent to a positive multiple of the extreme Reissner-Nordstr\"om solution.

Now for \textit{static} spacetimes, the normal Killing field $n$ is hypersurface orthogonal i.e. $n \wedge \td n =0$ everywhere. To first order in Gaussian null coordinates this is equivalent to\cite{Figueras:2011va} 
\be 
{\td} h^{(1)} = h^{(0)} \wedge h^{(1)} \qquad \mathrm{and} \qquad {\td} F^{(1)} = 2 F^{(1)} h^{(0)} \; . 
\ee
Therefore, we will call a transverse deformation static if it obeys these equations. For the case at hand these simplify to
\be 
{\td} h^{(1)} = 0  \qquad \mathrm{and} \qquad   {\td} F^{(1)} =  0 \; . 
\ee
With the general solution \eqref{RNsoln1},  we find that (\ref{E-a}) and (\ref{E+-}) imply
\bea
h^{(1)}= \frac{K_3 x - g(x)}{2 r_+^2} \td x+  \frac{2 K_1 x (1-x^2)}{ r_+^2} \td \phi \; , \qquad F^{(1)} = \frac{3 K_2 + K_3}{3 r_+^4} \; .
\eea 
Therefore our deformation is static if and only if $K_1 =0 $.  We deduce that there is a 2-parameter  family $(K_2, K_3)$ of static transverse deformations.  What do these correspond to?  

 It is useful to note that in order to compute the parameters $K_2, K_3$ that specify a transverse deformation of a known static solution, it is sufficient to compute $Q_2$  which only depends on $\gamma^{(1)}_{ab}$.  In the Appendix we give a general method for doing this. 

There are, of course, several known static and axisymmetric solutions to Einstein-Maxwell theory with extreme horizons. 
The most obvious  is the Majumdar-Papapetrou multi-black hole solution. The near-horizon geometry of each horizon is AdS$_2 \times S^2$. In the Appendix we compute the corresponding  first order deformation and we find that it is indistinguishable from the Reissner-Nordstr\"om solution, i.e., $K_2>0$ and $K_3=0$. Thus, perhaps surprisingly, multi-black holes are not visible at first order in the GNC; in other words, the extrinsic curvature of the horizon is unchanged by the presence of another black hole.

Another well known solution is the Reissner-Nordstr\"om solution in an external electric or magnetic field (a Melvin universe). However, there are no {\it static} solutions which are smooth in this family. Both the electric Reissner-Nordstr\"om in the electric Melvin background and the magnetic  Reissner-Nordstr\"om in the magnetic Melvin background suffer from conical singularities. The conical singularities may be avoided for e.g. an electric  Reissner-Nordstr\"om in a magnetic Melvin background (or vice-versa), however this solution is no longer static and is a special case of the general Kerr-Newman-Melvin solution (discussed in the next section).

So what is the interpretation of the parameter $K_3$?
In fact there is another smooth static and axisymmetric spacetime in Einstein-Maxwell theory with a extreme horizon\footnote{We thank Gary Gibbons for pointing out this solution.}: the extreme Ernst solution is a two-parameter family $(e, b)$ with $e>0$ and $e^2b^2<4$ corresponding to a static accelerating black hole with electric (or magnetic) charge $e$ held in equilibrium by a uniform electric (or magnetic) field $b$ (see Appendix) . Its near-horizon geometry is AdS$_2 \times S^2$~\cite{Dowker:1994up}, as must be the case from the near-horizon uniqueness theorem, with 
\be
r_+= \frac{e (4+ b^2 e^2)^2}{ 4( 4- b^2 e^2)}  \; .
\ee
Computing the first order transverse deformation and the corresponding gauge invariant variables (see Appendix), we find it takes the form of our general solution with
\be
K_1=0, \qquad K_2=\frac{ 2 e (4+ b^2 e^2)^3}{(4- b^2 e^2)^3}, \qquad K_3 = \frac{12 b^2 e^3( 4+b^2e^2)^3}{(4- b^2 e^2)^4} \; .
\ee
Thus the deformation in this case has both $K_2$ and $K_3$ nonvanishing.
For $b\to 0$ this reduces to the data for the  Reissner-Nordstr\"om black hole, as it should.  The scale invariant combination
\be
\frac{K_2+ \frac{K_3}{3}}{K_2}  =   \frac{ 4+b^2 e^2}{4- b^2 e^2} \geq 1  \; ,
\ee
where the inequality follows from the fact the RHS is a monotonic function in the domain $e^2 b^2<4$. This shows that not all marginally trapped deformations with $K_2\neq 0$ correspond to an Ernst solution. It would be interesting to determine if there are any black hole solutions which occupy the remaining part of the moduli space.

What about deformations with $K_2=0$? In fact, the extreme Kerr-Melvin solution admits a special case in which the near-horizon geometry is static (see Appendix). The deformation corresponding to this has $K_1=K_2=0$ and $K_3>0$. Therefore, although the extreme Kerr-Melvin solution is stationary and never static, in this special case it gives rise to a static deformation of AdS$_2\times S^2$ with $K_2=0$. For this solution staticity is broken at second order in the GNC expansion. Conversely, any marginally trapped, static deformation with $K_2=0$ corresponds to such a Kerr-Melvin solution.

Therefore, we have established Theorem \ref{thm:static} stated in the Introduction.
It is worth emphasising that above we also found a one-parameter family ($K_1$) of deformations which do not preserve staticity. We will discuss their interpretation in the next section where we analyse deformations of the more general extreme Kerr-Newman horizon.

\subsection{Kerr-Newman horizon}

The horizon data in this case is~\cite{Kunduri:2008tk} (see also Appendix)
\bea
\gamma^{(0)} &=& \frac{\rho_+^2  \td x^2}{1-x^2} + \frac{\left(1-x^2\right) \left(a^2+r_+^2\right)^2\td \phi^2}{\rho_+^2}   \\
h^{(0)} &=& -\frac{2 a^2 x}{\rho_+^2} \td x + \frac{2 a r_+ \left(1-x^2\right) \left(a^2+r_+^2\right)}{ \rho_+^4}  \td \phi \nn \\
 F^{(0)} &=& - \frac{a^4 x^4 +a^2 r_+^2 \left(6 x^2-4\right)+r_+^4 }{\rho_+^6}  \; , \nn  \\
 \Psi^{(0)} &=& \frac{a^2q_e x^2-2 aq_m r_+ x-q_e r_+^2}{\rho_+^4}\nn  \; , \\
 B^{(0)} & =& -\frac{\left(a^2+r_+^2\right) \left(a^2 q_m x^2+2 a q_e r_+ x-q_m r_+^2\right)}{\rho_+^4} \td x \wedge \td \phi  \nn  \; ,
\eea
where $\rho_+^2= r_+^2+ a^2 x^2$ and $r_+^2= a^2+ q_m^2+q_e^2$. We will assume $a>0$ and $q_m^2+q_e^2>0$. The horizon metric is defined for $-1<x<1$, $\phi$ is $2\pi$ periodic, and extends to a smooth metric on $S^2$ where the endpoints $x=\pm 1$ correspond to the fixed points of the axial symmetry.  

We will assume the first order deformation is also invariant under the axial Killing field $\partial_\phi$, in which case the gauge transformation function $f$ must also be axisymmetric~\cite{Li:2015wsa}. The gauge transformation rules (\ref{gautransgamma}) and (\ref{gautransZ}) are
\bea
&&\gamma^{(1)}_{xx} \to \gamma^{(1)}_{xx}+ f''(x) - \frac{x \left(a^2 \left(2 x^2-1\right)+r_+^2\right) f'(x)}{\left(1-x^2\right)\rho_+^2} \label{KNGT} \\
&&\gamma^{(1)}_{x\phi}\to \gamma^{(1)}_{x\phi} - \frac{a r_+ \left(1- x^2\right) \left(a^2+r_+^2\right) f'(x)}{\rho_+^4} \nn  \\
&&\gamma^{(1)}_{\phi\phi} \to \gamma^{(1)}_{\phi \phi} - \frac{x \left(1- x^2\right) \left(a^2+r_+^2\right)^3 f'(x)}{\rho_+^6}  \nn \\
&&Z^{(1)}_x \to Z^{(1)}_x -\frac{ \left(q_e \left(r_+^2-a^2 x^2\right)+2 a q_m  r x\right)f'(x)}{2 \rho_+^4 }  \nn \\  \nn 
&& Z^{(1)}_\phi \to Z^{(1)}_\phi - \frac{\left(1- x^2\right) \left(a^2+r_+^2\right)  \left(a^2 q_m x^2+2 a q_e r_+ x- q_m r_+^2\right)f'(x)}{2 \rho_+^6}  \; .
\eea
It is straightforward to check the following variables are invariant under the above gauge transformation:
\bea 
\nn Q_1 &=&  (a^2 +r_+^2) (1-x^2) \left[ - 2 a q_e r_+ x + q_m (r_+^2 - a^2 x^2) \right] Z^{(1)}_x \\
\nn && +(r_+^2 + a^2 x^2) \left[ 2 a q_m r_+ x + q_e(r_+^2 - a^2 x^2) \right] Z^{(1)}_\phi \\
\nn Q_2 &=& (a^2 + r_+^2)^3 x^2 (1-x^2)^2 \gamma^{(1)}_{xx} + x (1-x^2) (r_+^2 + a^2 x^2) ^3 {\gamma^{(1)'}_{\phi \phi}} \\
\nn && - (r_+^2 + a^2 x^2)^2 \left[ r_+^2 (1- 2x^2) + a^2 x^2 (5 x^2 -6) \right] \gamma^{(1)}_{\phi \phi}   \\
\nn Q_3 &=& - (q_e^2 + q_m^2) (r_+^2 + a^2 x^2)^2 \gamma^{(1)}_{x \phi} + 2 a r_+ (a^2 + r_+^2 ) (1-x^2) \left[ 2 a q_m r_+ x + q_e (r_+^2 - a^2 x^2) \right] Z^{(1)}_x \\
\nn && + 2 a r_+ (r_+^2 + a^2 x^2) \left[ 2 a q_e r_+ x + q_m (a^2 x^2 - r_+^2) \right] Z^{(1)}_\phi \\
Q_4 &=& (a^2 + r_+^2)^2 x \; \gamma^{(1)}_{x \phi} - a r_+ (r_+^2 + a^2 x^2 ) \gamma^{(1)}_{\phi \phi}  \label{KNgiquant} \; .
\eea
Note that these do not reduce to the variables used in the static case \eqref{RNgiquant} in the static limit $a\to 0$. However, they are closely related: in particular, as $a \to 0$ we have $Q_1 \to r_+^4 Q_1^{\text{static}}, Q_2 \to r_+^6 Q_2^{\text{static}}$. Nevertheless, as in the static case one can show these $Q_i$ are also globally defined and vanish at the poles of $S^2$.

In terms of these gauge invariant variables, the Maxwell equations \eqref{maxwell} are equivalent to (by taking linear combinations of the $x$ and $\phi$ components)
\bea
\nn 0 &=& (a^2 + r_+^2)^2 (1-x^2)(r_+^2 + a^2 x^2)^2 Q''_1 - 2 a^2 (a^2 + r_+^2)^2 x (1-x^2) (r_+^2 + a^2 x^2) Q'_1 \\
\nn && + 2 (a^2 + r_+^2)^2 \left[ - r_+^4 + 3 a^4 x^2 + a^2 r_+^2 (1 + x^2) \right] Q_1 + (a^2 + r_+^2)^2 (1-x^2) (r_+^2 + a^2 x^2) Q'_3 \\
&& - 2 a^2 (a^2 + r_+^2)^2 x (1-x^2) Q_3 +2 (r_+^4 -a^4)(r_+^2 + a^2 x^2)^2  Q_4  \label{KNME1}\\
\nn 0 &=& (a^2 - r_+^2 ) x^2(1-x^2)^2(r_+^2 + a^2 x^2)^4 Q''_4 + 2 (a^2 - r_+^2 ) x (1-x^2)^2 (r_+^2 + a^2 x^2)^3 (2 a^2 x^2 - r_+^2) Q'_4 \\
\nn && + 2 (a^2 - r_+^2 ) (1-x^2)(r_+^2 + a^2 x^2)^2 \left( r_+^4 - 2 r_+^4 x^2 + 3 a^4 x^4 + 2 a^2 r_+^2 x^4 \right) Q_4 \\
\nn && - (a^2 + r_+^2)^2 x^3 (1-x^2)^2 (r_+^2 + a^2 x^2)^2 Q''_3 + 2 a^2  
(a^2 + r_+^2)^2 x^4 (1-x^2)^2 (r_+^2 + a^2 x^2) Q'_3 \\
\nn && - 2 (a^2 + r_+^2)^2 x^3 (1-x^2) (- r_+^4 + 3 a^4 x^2 + a^2 r_+^2+ a^2 r_+^2 x^2 ) Q_3 \\
\nn && + a r_+ (a^2 - r_+^2) x (1-x^2)(r_+^2 + a^2 x^2)^2 Q'_2 + 2 a r_+  (a^2 - r_+^2) (3x^2 -1) (r_+^2 + a^2 x^2)^2 Q_2  \\
&& + 4 a^2 r_+^2  (a^2 + r_+^2)^2 x^3 (1-x^2)^2 (r_+^2 + a^2 x^2) Q'_1 - 
8 a^4 r_+^2  (a^2 + r_+^2)^2 x^4 (1-x^2)^2 Q_1 \label{KNME2} \; .
\eea
The $x\phi$ component of the Einstein equation \eqref{einstein} gives
\bea 
\nn 0 &=& x^2 (1-x^2)^2 (r_+^2 + a^2 x^2)^3 Q''_4 - 2 x (1-x^2) (r_+^2 + a^2 x^2)^2 (r_+^2 - a^2 x^2 + 2 a^2 x^4) Q'_4 \\
\nn && - 2 (r_+^2 + a^2 x^2) \left[ 2 a^2 r_+^2 x^4 (1+x^2) + (r_+^4+ a^4 x^4) (3 x^2 -1)\right]Q_4 + a r_+ x (1-x^2) (r_+^2 + a^2 x^2) Q'_2 \\
\nn && 2 a r_+ (3x^2 -1) (r_+^2 + a^2 x^2)  Q_2 + 4 (a^2 + r_+^2)^2 x^3 (1-x^2) (r_+^2 + a^2 x^2) Q'_1 \\
&& + 8 (a^2 + r_+^2)^2 x^4 (r_+^2 + a^2 x^2) Q_1 - 4 (a^2 + r_+^2)^2 x^3 (1-x^2) Q_3 \label{KNEE1} \; , 
\eea
and the $x\phi$ component plus $(a^2 +r_+^2)$ times the $xx$ component of the Einstein equation gives
\bea
\nn 0 &=& (1-x^2)^2 (r_+^2 + a^2 x^2)^3 Q''_4 - 2 x (1-x^2) (r_+^2 + a^2 x^2)^3 Q'_4 \\
\nn && - 2 (r_+^2 + a^2 x^2)^2 \left[ r_+^2(3 x^2 -1) + 2 a^2 - a^2 x^2(1-x^2)\right] Q_4 + 2 (a^2 + r_+^2) (1-x^2)^2 (r_+^2 + a^2 x^2)Q'_3 \\
\nn && - 4 a^2 (a^2 + r_+^2) x (1-x^2)^2 Q_3 + 4 (a^2 + r_+^2)^2 x (1-x^2)  (r_+^2 + a^2 x^2)Q'_1\\
&& + 8 (a^2 + r_+^2) \left[ x^2 (r_+^4 + a^4 x^2) + a^2 r_+^2 + a^2 r_+^2 x^2(2x^2-1) \right]Q_1 \label{KNEE2} \; . 
\eea
The remaining components of the Einstein equation are automatically satisfied due to the fact it is traceless.

The Maxwell equation \eqref{KNME2} can be further simplified by subtracting from it $(a^2 - r_+^2)  (r_+^2 + a^2 x^2)$ times the Einstein equation \eqref{KNEE1} resulting in
\bea 
\nn 0 &=& - (a^2 + r_+^2)^2 (1-x^2)^2   (r_+^2 + a^2 x^2)^2 Q''_3 + 2 a^2 (a^2 + r_+^2)^2 x (1-x^2)^2   (r_+^2 + a^2 x^2) Q'_3 \\
\nn && - 2 (a^2 + r_+^2)^2 (1-x^2) (r_+^4 + a^2 x^2 - a^2 r_+^2 + 3 a^2 r_+^2 x^2) Q_3 + 2 (a^4 -r_+^4)(1-x^2)(r_+^2 + a^2 x^2)^3 Q'_4 \\
\nn && + 4 (a^2 - r_+^2)(a^2 + r_+^2)^2 x (r_+^2 + a^2 x^2)^2 Q_4 + 4 (a^2 + r_+^2)^2 (1-x^2) (r_+^2 + a^2 x^2) (r_+^4 - a^4 x^2) Q'_1 \\
&& -8 (a^2 + r_+^2)^2  x (a^4 r_+^2 - r_+^6 + a^6 x^4 + a^2 r_+^4 - 2 a^2 r_+^4 x^2 ) Q_1 \label{KNeqn1} \; . 
\eea
Therefore, the Einstein-Maxwell equations are equivalent to the four equations \eqref{KNME1}, \eqref{KNEE1}, \eqref{KNEE2}, \eqref{KNeqn1}; note only \eqref{KNEE1} contains $Q_2$ terms, so let us focus on the remaining three equations first.

We can rearrange \eqref{KNME1} and write 
\bea
\nn (r_+^2 + a^2 x^2) Q'_3  - 2 a^2 x \; Q_3  &=& \frac{1}{(a^2 + r_+^2)^2(1-x^2)} \left[ -(a^2 + r_+^2)^2(1-x^2)(r_+^2 + a^2 x^2)^2 Q''_1 \right. \\
\nn &&  + 2 a^2 (a^2 + r_+^2)^2 x (1-x^2)(r_+^2 + a^2 x^2)Q'_1 \\
\nn && - 2 (a^2 + r_+^2)^2 (- r_+^4 + 3 a^4 x^2 +a^2 r_+^2 +a^2 r_+^2  x^2) Q_1  \\
&& \left. + 2 (a^4 - r_+^4) (r_+^2 + a^2 x^2)^2 Q_4 \right] \label{KNeqn2} \; , 
\eea
and express \eqref{KNeqn1} as 
\bea
\nn 0 &=& -(a^2 + r_+^2)^2 (1-x^2) \left\lbrace 
 (1-x^2)(r_+^2 + a^2 x^2) \left[(r_+^2 + a^2 x^2) Q'_3  - 2 a^2 x \; Q_3 \right]' \right. \\
&& \left.- 2 a^2 x (1-x^2) \left[(r_+^2 + a^2 x^2) Q'_3  - 2 a^2 x \; Q_3 \right] + 2 (r_+^2 + a^2 x^2)^2 Q_3 \vphantom{\sum} \right\rbrace + ... \; ,  \label{KNeqn3}
\eea
where $...$ denote the terms not involving $Q_3$.  Thus, using  \eqref{KNeqn2} we can substitute for $(r_+^2 + a^2 x^2) Q'_3  - 2 a^2 x \; Q_3 $ in \eqref{KNeqn3} to find an expression for $Q_3$ in terms of $Q_1$ and $Q_4$ terms. It turns out that the $Q_4$ terms all cancel and we are left with simply 
\be
2 (1-x^2) Q_3 =  (1-x^2)^2 (r_+^2 + a^2 x^2) Q'''_1 + 2 (1-x^2)(r_+^2 + a^2 x^2) Q'_1 + 4 (a^2 + r_+^2) x \; Q_1  \label{KNQ3eqn} \; . 
\ee
Now, using \eqref{KNQ3eqn} we can substitute for all the $Q_3$ terms in \eqref{KNME1} to get 
\bea
\nn 0 &=& 4 (a^2 -r_+^2) (1-x^2)  (r_+^2 + a^2 x^2) Q_4 - (a^2 +r_+^2) (1-x^2)^3 (r_+^2 + a^2 x^2) Q^{(4)}_1 \\
\nn && +2 (a^2 +r_+^2) x(1-x^2)^2  (r_+^2 + a^2 x^2) Q'''_1 - 4 (a^2 +r_+^2) (1-x^2)^2  (r_+^2 + a^2 x^2) Q''_1 \\
&& - 4 (a^2 +r_+^2) x (1-x^2) (r_+^2 + a^2 x^2) Q'_1 -8 (a^2 +r_+^2)  (r_+^2 + a^2 x^2) Q_1  \label{KNQ4eqn} \; , 
\eea
which allows us to solve for $Q_4$ in terms of $Q_1$ (note $r_+^2-a^2= q_e^2+q_m^2>0$ by assumption).

We may obtain another equation for $Q_1$ and $Q_4$ only by taking the linear combination $-2(1-x^2)$ \eqref{KNME1} $+ (a^2+ r_+^2)$ \eqref{KNEE2} to eliminate the $Q_3$ terms, resulting in
\bea
\nn 0 &=& -2 (a^2+ r_+^2)(1-x^2)^2 (r_+^2 + a^2 x^2)^2 Q''_1 \\
\nn &&+ 4 (a^2+ r_+^2) x (1-x^2)(r_+^2 + a^2 x^2)(r_+^2 + 2 a^2 - a^2 x^2) Q'_1 \\
\nn && + 4  (a^2+ r_+^2)^2 (r_+^2 + r_+^2 x^2 - 3 a^2 x^2 + 5 a^2 x^4)Q_1  +  (1-x^2)^2 (r_+^2 + a^2 x^2)^3 Q''_4
\\
&&- 2 x (1-x^2) (r_+^2 + a^2 x^2)^3 Q'_4  - 2 (1-x^2) (r_+^2 + a^2 x^2)^3 Q_4 \label{KNeqn4}  \; . 
\eea
Eliminating $Q_4$ from the equations \eqref{KNQ4eqn} and \eqref{KNeqn4} then gives us a remarkably simple sixth order equation for $Q_1$:
\be
0 = (1-x^2)^2 Q_1^{(6)} - 12 x (1-x^2) Q_1^{(5)} - 6 (1- 5 x^2) Q_1^{(4)} \label{KNQ1eqn} \; .
\ee
The general solution to \eqref{KNQ1eqn} which is regular at $x= \pm 1$ is simply the polynomial
\be 
Q_1= A_1 x^3 +\tilde{A}_2 x^2 +\tilde{A}_3 x +A_4 
\ee
where $A_1, \tilde{A}_2, \tilde{A}_3, A_4$ are constants
(we have discarded the log terms in the most general solution to the differential equation \eqref{KNQ1eqn} since they are not regular at the poles). 

We can use \eqref{KNQ4eqn} to determine the general solution to $Q_4$. Regularity at the poles forces $\tilde{A}_3 = - A_1$ and $A_4 = - \tilde{A}_2$, thus  giving
\be 
Q_1 = - (\tilde{A}_2 + A_1 x) (1-x^2) \; , \qquad Q_4= -\frac{2 (a^2 + r_+^2)(\tilde{A}_2 + A_1 x) (1-x^2)}{r_+^2 + a^2 x^2} \; , 
\ee 
and from \eqref{KNQ3eqn} we deduce
\be 
Q_3= 2 (A_1 r_+^2  - \tilde{A}_2 a^2  x)(1-x^2) \; .
\ee
Finally, from \eqref{KNEE1} we find
\be 
Q_2 = \frac{8 A_1 a (a^2 + r_+^2) x^3 (1-x^2)^2}{r_+} - \frac{2 \tilde{A}_2 r_+ (a^2 + r_+^2) (1-x^2)^2}{a} + A_3 x^2(1-x^2)^2 \; , 
\ee
where $A_3$ is an integration constant. In order to able to take the $a \rightarrow 0$ static limit to  compare with our analysis of the static solutions, without loss of generality we will define the constant $A_2 = \tilde{A}_2/a$. We have therefore found the general solution for the gauge invariant variables $Q_1, Q_2, Q_3, Q_4$ parameterised in terms of constants $A_1, A_2, A_3$.

Now we can invert the equation \eqref{KNgiquant} to get the general solution for the individual components of $\gamma^{(1)}_{ab}$ and $Z^{(1)}_a$. Let us choose $\gamma^{(1)}_{\phi \phi}$ to be the arbitrary function corresponding to the gauge freedom. We may write 
\be 
\gamma^{(1)}_{\phi \phi} = \frac{(1-x^2)\left( m + x h(x)\right)}{(r_+^2 + a^2 x^2)^2} 
\ee
where $m$ is some constant and $h(x)$ is some smooth function. Under a gauge transformation (\ref{KNGT}) we find $h(x)\to h(x) - (a^2+r_+^2)^3 f'(x)/\rho_+^2$, so the function $h(x)$ may be thought of as parameterising the gauge freedom. The solution for $Q_4$ implies
\bea
\gamma^{(1)}_{x \phi} &=& -\frac{2 A_1 (1-x^2)}{(a^2+r_+^2)(r_+^2 + a^2 x^2)} + \frac{a m r_+ (1-x^2) - 2 A_2 a (a^2+r_+^2) (1-x^2) }{(a^2+r_+^2)^2 x (r_+^2 + a^2 x^2)} \nn \\ &&+ \frac{a r_+ (1-x^2)h(x) }{(a^2+r_+^2)^2  (r_+^2 + a^2 x^2)} \; . 
\eea
To avoid the pole at $x=0$ we must set
\be 
m = \frac{2 A_2 (a^2+r_+^2)}{r_+} \; . 
\ee
In summary, the general solution is a three parameter family parameterised by constants $A_1, A_2, A_3$, and a smooth function $h(x)$ corresponding to the gauge freedom, where
\bea
\nn \gamma^{(1)}_{xx} &=& \frac{8 A_1 a x}{r_+ (a^2+r_+^2)^2} + \frac{2 A_2 (r_+^2 - 2 a^2 + 3 a^2 x^2)}{r_+ (a^2+r_+^2)^2 (1-x^2)} + \frac{A_3}{(a^2+r_+^2)^3} \\
\nn && - \frac{ (1-x^2) (r_+^2 + a^2 x^2) h'(x) - x (r_+^2 - 3 a^2 + 4 a^2 x^2)h(x)}{(a^2+r_+^2)^3(1-x^2)} \\
\nn \gamma^{(1)}_{x \phi} &=&  \frac{a r_+ h(x) - 2 A_1 (a^2+r_+^2)}{(a^2+r_+^2)^2 (r_+^2 + a^2 x^2)} (1-x^2)\\
\nn \gamma^{(1)}_{\phi \phi} &=& \frac{2 A_2 (a^2+r_+^2) + r_+ x h(x)}{r_+(r_+^2 + a^2 x^2)^2 }(1-x^2) \\
\nn Z^{(1)}_x &=&  A_1 \frac{q_e(2 a r_+^2 + a r_+^2 x^2-a^3 x^2) + q_m (3 a^2 r_+ x - r_+^3 x)}{r_+ (r_+^2-a^2) (a^2+r_+^2)^2(r_+^2 + a^2 x^2)} 
+ A_2 \frac{a (a q_e x - q_m r_+)}{r_+ (r_+^4-a^4)(r_+^2 + a^2 x^2)} \\
\nn && + \frac{2 a q_m r_+ x + q_e (r_+^2 - a^2 x^2)}{2(a^2+r_+^2)^3(r_+^2 + a^2 x^2)}h(x) \\
\nn Z^{(1)}_\phi &=&  -A_1 \frac{q_m(2 a r_+^2 + a r_+^2 x^2-a^3 x^2) - q_e (3 a^2 r_+ x - r_+^3 x)}{r_+ (r_+^4-a^4) (r_+^2 + a^2 x^2)^2}(1-x^2) \\
 \nn && - A_2 \frac{a (a q_m x + q_e r_+)}{r_+ (r_+^2-a^2)(r_+^2 + a^2 x^2)^2} (1-x^2) + \frac{2 a q_e r_+ x - q_m (r_+^2 - a^2 x^2)}{2(a^2+r_+^2)^2(r_+^2 + a^2 x^2)^2} (1-x^2)h(x)\; .  \\ 
\eea
It can be checked that near the poles $x^2 \rightarrow 1$ we have 
\be
\gamma^{(1)}_{xx} = \gamma^{(1)}_{\phi \phi} (1-x^2)^{-2} + \mathcal{O}(1) \qquad \gamma^{(1)}_{x\phi}=O(1-x^2) \qquad \gamma^{(1)}_{\phi\phi}=O(1-x^2)
\ee
 so the first order quantity $\gamma^{(1)}_{ab}$ is a indeed smooth tensor field on $S^2$~\cite{Li:2015wsa}.   In the static limit $a\to 0$ our solution reduces to the general static solution with $K_1= - A_1/r_+^4, K_2 =2 A_2/ r_+^3, K_3= A_3 /r_+^6$.

We will now impose that the deformation is such that $S$ is a MTS.  The appropriate gauge invariant quantity is  the mean expansion $\int_S \Gamma \gamma^{(1)}$ where  $\Gamma>0$ is the unique (up to scale) function such that that $h^{(0)} = \Gamma^{-1} \tilde{h}^{(0)} - \td \log \Gamma$ and $D \cdot \tilde{h}^{(0)} =0$~\cite{Li:2015wsa}. For the extreme Kerr-Newman horizon one can set $\Gamma= \rho_+^2/(r_+^2+a^2)$~\cite{Kunduri:2008tk} and we find 
\be 
\int_S \Gamma \gamma^{(1)}  = \frac{8 \pi}{3(a^2 + r_+^2)^3}  (A_3 + 6 A_2 r_+ (a^2 + r_+^2 )) \; . 
\ee
Therefore, the MTS condition for our first order deformation requires 
\be
A_3 + 6 A_2 r_+ (a^2 + r_+^2 )>0  \; .
\ee
The extreme Kerr-Newman black hole solution gives rise to a transverse deformation (see Appendix). It is easily checked this takes the form of our solution with 
\be
A_1=0, \qquad A_2 = (q_e^2 + q_m^2)(a^2 + r_+^2), \qquad A_3 = 12 a^2 r_+ (a^2 + r_+^2)^2,
\ee
which clearly satisfies our MTS condition.  It is worth noting that in order to extract the parameters $A_1, A_2, A_3$ of a transverse deformation for a given known solution, it is sufficient to compute $Q_2$ which in fact only depends on $\gamma^{(1)}_{ab}$. We give a general recipe for doing this in the Appendix.

Another family of stationary and axisymmetric solutions with an extreme horizon is given by the Kerr-Newman-Melvin solution, i.e. a Kerr-Newman black hole in an external electric/magnetic field.  This is a rather complicated solution so it is helpful to consider a few instructive special cases.  

The extreme Kerr-Melvin solution is a 2-parameter family parameterised by $(\tilde{a}, b)$, corresponding to the rotation parameter $\tilde{a}$ and the external magnetic field $b$ (see Appendix).  It's near-horizon geometry is isometric to that of an extreme Kerr-Newman black hole with parameters
\be
r^2_+= \tilde{a}^2(1+ \tilde{a}^2 b^2)^2, \qquad a^2= \tilde{a}^2 (1- \tilde{a}^2 b^2)^2  \; .
\ee
A computation reveals that the first order transverse deformation then takes our general form (see Appendix) with
\be
A_1=A_2=0, \qquad A_3 = 48 \tilde{a}^7 (1+ \tilde{a}^4 b^4)^3  \; , 
\ee
which clearly satisfies our MTS condition.
Notice this is of a different form to the deformation corresponding to the Kerr-Newman black hole (which has $A_2 \neq 0$); thus at first order it is possible to distinguish an external magnetic field from intrinsic charge.  

Another interesting special case is the extreme Reissner-Nordstr\"om-Melvin solution which is also a 2-parameter family parameterised by $(\tilde{q}, b)$, corresponding to the electric charge parameter $\tilde{q}$ and external magnetic field $b$ (see Appendix). Its near-horizon geometry is again given by  that of an extreme Kerr-Newman black hole with parameters
\be
r^2_+ = \tilde{q}^2 (1+ \tfrac{1}{4} \tilde{q}^2 b^2)^2, \qquad a^2 = \tilde{q}^4 b^2  \; ,
\ee
and the corresponding first order deformation is given by
\bea
A_1&=&0,  \qquad A_2= \tfrac{1}{64} \tilde{q}^3 (4-b^2 \tilde{q}^2) (b^4 \tilde{q}^4+24 b^2 \tilde{q}^2+16), \nn  \\  A_3 &=& \tfrac{3}{1024} b^2 \tilde{q}^8 (b^2 \tilde{q}^2+12) (b^4 \tilde{q}^4+24 b^2 \tilde{q}^2+16)^2  \; .  \label{ARNM}
\eea
It may be verified that this obeys the MTS condition.

The general extreme Kerr-Newman-Melvin is a much more complicated solution parameterised by three-parameters $(\tilde{a},\tilde{q}, b)$. It's near-horizon geometry has parameters~\cite{Booth:2015nwa, Hejda:2015gna, Bicak:2015lxa}
\be
r_+= \tilde{r}_+ + \tilde{a} \tilde{q} b + \tfrac{1}{4} \tilde{r}_+ (3 \tilde{a}^2+\tilde{r}_+^2)b^2, \qquad a= \tilde{a} - \tilde{q} \tilde{r}_+ b - \tfrac{1}{4} \tilde{a} (\tilde{a}^2+3 \tilde{r}_+^2)b^2
\ee
where $\tilde{r}_+= \sqrt{\tilde{a}^2+\tilde{q}^2}$.
We have verified that the corresponding deformation also has $A_1=0$ and generically $A_2, A_3 \neq 0$; indeed, this case interpolates between Kerr-Newman, Kerr-Melvin and Reissner-Nordstr\"om-Melvin described above. It is worth noting that although the extreme Kerr-Newman-Melvin solution is a stationary and axisymmetric solution, it includes a two parameter special case for which the near-horizon geometry is static and hence isometric to the AdS$_2\times S^2$ solution. As noted above, in the static horizon limit $a\to 0$ we have $A_1\to - r_+^4 K_1$, so we deduce that this special case of Kerr-Newman-Melvin solutions can only lead to deformations of AdS$_2\times S^2$ with $K_1=0$, i.e. they preserve staticity to first order in GNC even though the full solution is  not static.

All known solutions discussed so far have $A_1=0$. So what about the interpretation of the parameter $A_1$?
In fact, as in the static case, a more general class of spacetimes containing a smooth extremal horizon maybe be constructed.  By applying a Harrison transformation to a rotating, dyonic C-metric one can construct a regular solution which corresponds to a rotating, dyonic, accelerating black hole held in equilibrium by a uniform magnetic field~\cite{Astorino:2016xiy}. This is a rotating version of the Ernst solution, so we will simply refer to it as the rotating Ernst solution.  It has an extremal limit where the surface gravity of the horizon is zero. The resulting extreme solution is a four parameter family parameterised by $(\tilde{a}, \tilde{q}, \tilde{p} , A, b)$, subject to a constraint coming from the removal of conical singularities at $x = \pm 1$.  Here, $(\tilde{a}, \tilde{q}, \tilde{p})$ are the rotation parameter, electric and magnetic charges, $A$ is the acceleration parameter, and $b$ is the external magnetic field.  

For simplicity we will only consider the special case $\tilde{a}=0$ given in~\cite{Astorino:2016xiy}; for $\tilde{q} \neq 0$ even this case is rotating due to the external magnetic field and hence sufficiently general for our purposes. As shown in the Appendix, the regular solution can be parameterised by $z=\sqrt{\tilde{q}^2+\tilde{p}^2}$ and $(A,b)$ subject to certain inequalities. 
Its near-horizon geometry is isometric to that of the Kerr-Newman black hole, as it must be by the near-horizon uniqueness theorems, with parameters $(r_+, a)$ given by (\ref{rotErnstparam}).  The corresponding first order deformation takes our general form with $A_1, A_2, A_3$ given by (\ref{rotErnstAi}); in particular note that generically all three constants $A_1, A_2, A_3$ are nonvanishing. For $A=0$, this solution reduces to Reissner-Nordstr\"om-Melvin, in which case $A_1=0$ as found above. This provides an interpretation for the parameter $A_1$. However, as in the static case, presumably the rotating Ernst solution does not occupy all parts of the moduli space, i.e. although this gives a three parameter family of deformations not all values of $A_1, A_2, A_3$ may be  realised. It would be interesting if there are new solutions which fill out the rest of the moduli space, or if there is some obstruction to promoting our linearised solutions to higher order in the GNC expansion in these regions of moduli space. 

We have now established Theorem \ref{thm:KN} stated in the Introduction. It is worth noting that this general four parameter family of extreme rotating Ernst solutions discussed above will contain a three parameter family of solutions with a static near-horizon geometry. Presumably the corresponding deformations will give the non-static $K_1\neq 0$ deformations of the AdS$_2\times S^2$ near-horizon geometry found in the previous section.

\section{Uniqueness of solutions with a cosmological constant}
\label{sec:ads}

Here we will consider the vacuum case with a cosmological constant $\Lambda$. The possible static near-horizon geometries with compact cross-sections $S$ are given by 
\be
F^{(0)}= \Lambda, \qquad  h^{(0)}_a=0, \qquad R^{(0)}_{ab}=\Lambda \gamma^{(0)}_{ab},  \label{statichorizon}
\ee
corresponding to $dS_2\times S^{2}$ or $AdS_2\times H^2$ if $\Lambda>0$ and $\Lambda<0$ respectively~\cite{Chrusciel:2005pa}.  The most general axisymmetric near-horizon geometry is given by that of Kerr-(A)dS~\cite{Kunduri:2008rs}. We will determine the moduli space of transverse deformations of these near-horizon geometries. 

\subsection{Static horizons}

In this case, the gauge freedom (\ref{gautransgamma}) reduces to $\gamma_{ab}^{(1)}\to \gamma_{ab}^{(1)}+ D_a D_b f$ and may be used to fix a gauge in which $\gamma^{(1)}$ is a constant. This  can always be done since it involves solving Poisson's equation on a compact manifold $S$. For the sake of generality, we will assume that $(S, \gamma_{ab}^{(0)})$ is an $n=D-2$ dimensional  maximally symmetric space, so
\be
R^{(0)}_{abcd}= K( \gamma^{(0)}_{ac} \gamma^{(0)}_{bd}- \gamma^{(0)}_{ad} \gamma^{(0)}_{bc}),
\ee
where $\Lambda = K (n-1)$. For $D=4,5$, so $S$ is 2 or 3 dimensional, this is the only possibility since $S$ is an Einstein space by equation (\ref{statichorizon}). The linearised Einstein equation (\ref{einstein}) reduces to
\be
- D^2 \tilde{\gamma}_{ab}^{(1)} = - 2n K \tilde{\gamma}^{(1)}_{ab}  \; ,
\ee
where ${\tilde{\gamma}}_{ab}^{(1)} = \gamma_{ab}^{(1)} - \frac{1}{n} \gamma^{(1)} \gamma^{(0)}_{ab}$ is the traceless part of the deformation.

First assume $\Lambda>0$, so $S$ is locally isometric to the round metric on $S^{n}$. Then, since $-D^2$ is a positive definite operator on a compact manifold, we must have ${\tilde{\gamma}}_{ab}^{(1)} \equiv 0$. Thus, the only deformations of $dS_2\times S^{n}$ are given by  $\gamma_{ab}^{(1)} = \frac{1}{n} \gamma^{(1)} \gamma^{(0)}_{ab}$. These correspond to the Narai solution (this is the extreme limit of Schwarzschild de-Sitter in which the black hole and cosmological horizons are coincident).

Now consider $\Lambda<0$, so $S$ is locally hyperbolic space $H^n$. The above argument no longer works and we have to work a little harder.  The Einstein equation states that ${\tilde{\gamma}}_{ab}^{(1)} $ is an eigentensor of $-D^2$ with eigenvalue $2n|K|$. First consider $n=2$, in which case $S\cong \Sigma_g$ where $\Sigma_g$ is a Riemann surface of genus $g \geq 2$. It is well known that in general $n=2$ symmetric tensor harmonics are all derived from scalar harmonics on $S$. A basis for traceless symmetric tensor harmonics is given by
\be
D_aD_b Y - \frac{1}{n} \gamma^{(0)}_{ab} D^2 Y, \qquad D_{(a} \epsilon_{b)}^{(0) c} D_c Y  \; ,   \label{tensorharmonics}
\ee
where $\epsilon_{ab}^{(0)}$ is the volume form of $(S, \gamma^{(0)}_{ab} )$ and $Y$ are the scalar harmonics obeying  $-D^2 Y = \lambda Y$. It can be checked that (\ref{tensorharmonics}) are eigentensors of $-D^2$ both with eigenvalue $\lambda - 2n K$. Comparing to the Einstein equation we deduce that $\tilde{\gamma}^{(1)}_{ab}$ must be a linear combination of the $\lambda=0$ harmonics. However, the only $\lambda=0$ scalar harmonics on a compact manifold are $Y=$constant, so in fact we must have ${\tilde{\gamma}}_{ab}^{(1)} \equiv 0$. Thus, we also deduce that the only deformations of AdS$_2\times \Sigma_g$ are  given by  $\gamma_{ab}^{(1)} = \frac{1}{n} \gamma^{(1)} \gamma^{(0)}_{ab}$. These correspond to the extreme Schwarzschild-AdS$_4$-hyperbolic black hole.  

This argument fails for $n>2$, since then one can have non-scalar derived tensor harmonics. Hence, the moduli space of deformations may be more complicated in this case, and need not correspond to the Schwarzschild-AdS-hyperbolic black hole. 

\subsection{Kerr-AdS horizon}

The extreme Kerr-AdS horizon data with $\Lambda = - 3g^2$ is given by~\cite{Kunduri:2008rs} (see also Appendix)
\bea
\gamma^{(0)} = \frac{\rho_+^2}{(1-x^2)\Delta_x} \td x^2  + \frac{(r_+^2+a^2)  (1-x^2)\Delta_x }{\rho_+^2 \Xi^2} \td \phi^2  \\
h^{(0)} = \frac{ 2a r_+ (r_+^2+a^2) (1-x^2)\Delta_x}{\Xi \rho_+^2} \td \phi   - \frac{2 a^2 x}{\rho_+^2} \td x  \; ,
\eea
where $\rho_+^2 = r_+^2+a^2 x^2$, $\Delta_x = 1- a^2 g^2 x^2$ and $\Xi  = 1 -a^2 g^2$ and the parameters obey $0<r_+<g^{-1}$ and
\be
a = r_+ \sqrt{ \frac{1+3 g^2 r_+^2}{1-g^2 r_+^2}}   \; .
\ee
The coordinate ranges are $-1<x<1$ and $\phi$ is $2\pi$-periodic and the endpoints $x= \pm 1$ are coordinate singularities corresponding to the fixed points of the axial symmetry (the poles of $S^2$).

We will consider axisymmetric deformations, in which case as noted earlier the function $f$ parameterising the allowed gauge transformation must also be axisymmetric. It is convenient to introduce gauge invariant variables.  It  is straightforward to check that
\be
X = (r_+^2+a^2) x [ -r_+^2+ a^4 g^2 x^4 + a^2 (-1+g^2 r_+^2(-1+2 x^2))] \gamma^{(1)}_{x \phi}(x) + a r_+ \rho_+^2 \Xi \gamma^{(1)}_{\phi\phi}
\ee
is invariant under our gauge transformations (\ref{gautransgamma}). Further, $X$ is smooth and vanishes at the poles $x=\pm 1$.  Note that for $g=0$ this is proportional to the gauge invariant variable (also called $X$) used in the pure vacuum case~\cite{Li:2015wsa}. One can then use this to eliminate $\gamma_{\phi\phi}^{(1)}$ in favour of $X$ and doing this for the $x\phi$ component of the linearised Einstein equation \eqref{einstein} gives
\bea
(1-x^2)X'  &+& \frac{2x}{5  \Delta_x \rho_+^2 } [ - 4r_+^2 + a^4 g^2 x^2 (-1+5 x^2) \nn \\ &&+ a^2 (-3-x^2+4 g^2 r_+^2 (-1+ 2x^2)] X + Y=0 \; ,  \label{Yeq}
\eea
where $Y$ is another gauge invariant variable which is a complicated linear combination of $\gamma^{(1)}_{x\phi}, {\gamma^{(1)}_{x\phi}}', {\gamma^{(1)}_{x\phi}}'', {\gamma^{(1)}_{xx}}'$ (its explicit form is unilluminating). Similarly, the $\phi\phi$ component of the linearised Einstein equation reduces to an equation of the form
\bea
a_1 X''+ a_2 X' + a_3 X + Z=0   \label{Zeq}
\eea
where $a_1, a_2, a_3$ are complicated functions of $x$ and $Z$ is another gauge invariant variable which is a linear combination of $\gamma^{(1)}_{x\phi}, {\gamma^{(1)}_{x\phi}}', {\gamma^{(1)}_{x\phi}}'', \gamma^{(1)}_{xx}, {\gamma^{(1)}_{xx}}'$.  Because the linearised Einstein equation is automatically traceless the final component is redundant.  

In the pure vacuum case $g=0$ one can check that $Z \propto Y$ 
so these two variables are not independent; one may then eliminate $Y$ to get a second order ODE for $X$ as found in~\cite{Li:2015wsa}. However, for $g \neq 0$ the variables $Y$ and $Z$ are independent and we must proceed differently, as follows.

For $g \neq 0$ one can invert the definitions of $Y$ and $Z$ to solve algebraically for $\gamma^{(1)}_{xx}, {\gamma^{(1)}_{xx}}'$ in terms of $Y, Z, \gamma^{(1)}_{x\phi}, {\gamma^{(1)}_{x\phi}}', {\gamma^{(1)}_{x\phi}}''$. Then imposing the `integrability condition' ${\gamma^{(1)}_{xx}}' = \td\gamma^{(1)}_{xx}/\td x$ results in a gauge invariant equation of the form
\be
b_1 Y' + c_1 Z'+ b_2 Y +c_2 Z=0  \label{inteq}  \; ,
\ee
where $b_1, b_2, c_1, c_2$ are unsightly functions of $x$.

Using (\ref{Yeq}) and (\ref{Zeq}) to eliminate $Y$ and $Z$ in (\ref{inteq}) finally gives  the following third order ODE for $X$,
\be
\alpha_1 X'''+ \alpha_2 X'+ \alpha_3 X=0
\ee
where
\bea
&& \alpha_1= \left(1-x^2\right)^3 \left(1- 3 g^4 r_+^4 x^2- g^2 r_+^2 \left(x^2+1\right)\right)^3   \\  && \alpha_2= -4 \left(1-3 g^2 r_+^2\right)^2 \left(1-x^2\right) \left(1- 3 g^4 r_+^4 x^2- g^2 r_+^2 \left(x^2+1\right)\right) \left\{3 g^4 r_+^4 x^4 \right. \nn \\ && \qquad \qquad  \left. +g^2 r_+^2 \left(x^4+6 x^2-1\right)+2 x^2+1\right\}  \\ && \alpha_3= 8 x \left(1-3 g^2 r_+^2\right)^2  \left\{  9 g^8 r_+^8 x^6+3 g^6 r_+^6 \left(2 x^6+9 x^4-4 x^2+1\right) \right. \nn \\ && \qquad \qquad \left. +g^4 r_+^4 x^2 \left(x^4+18 x^2-1\right)+g^2 r_+^2 \left(3 x^4-2 x^2-1\right)-x^2-2\right\}  \; .
\eea
Remarkably, one can find the general solution to this equation in terms of elementary functions.  The general solution which is smooth at $x \to \pm 1$ is simply
\be
X = -\frac{A (1-x^2) [ 4 - x^2( 1+ 3 g^2 r_+^2)^2]}{4 [ 1- 3 g^4 r_+^4 x^2 - g^2 r_+^2 (1+x^2)]}
\ee
where $A$ is an integration constant. The other functions $Y$ and $Z$ are then also determined by (\ref{Yeq}) and (\ref{Zeq}). Using $X,Y,Z$, one can then solve for the deformation $\gamma^{(1)}_{xx}, \gamma^{(1)}_{\phi\phi}$ algebraically in terms of  $\gamma^{(1)}_{x\phi}(x)$ (an arbitrary function reflecting the gauge freedom) and the constant $A$  (i.e. there are no further integration constants). 

Using this general solution one can then compute the gauge invariant mean expansion $\int_S \Gamma \gamma^{(1)}$. In this case we may take~\cite{Kunduri:2008rs} 
\be
\Gamma = \frac{\rho_+^2}{\Xi (r_+^2+a^2)}
\ee
and we find
\be
\int_S \Gamma \gamma^{(1)} = \frac{ \pi A}{2 g^2 r_+^6 (1+g^2 r_+^2)^3 (1- 3 g^2 r_+^2)} \sqrt{\frac{ (1-g^2 r_+^2)^7}{1+ 3 g^2 r_+^2}}
\ee
so the MTS condition is simply $A>0$.

Now, using the first order data for the extreme Kerr-AdS black hole (see Appendix), it is straightforward to verify that $X$ takes precisely the above form (as it must) with
\be
A = \frac{32 g^2 r_+^7 (1+g^2 r_+^2)}{(1- 3 g^2 r_+^2)} \sqrt{ \frac{ 1+ 3 g^2 r_+^2}{(1- g^2 r_+^2)^3} }
\ee
which indeed obeys the MTS condition.
Since such linear deformations are only determined up to scale, we deduce that the general solution must be gauge equivalent to the first order data of the extreme Kerr-AdS black hole.  This establishes Theorem \ref{thm:ads}.

Therefore, the uniqueness theorem established for the vacuum extreme Kerr horizon~\cite{Li:2015wsa}, persists with a cosmological constant.   We emphasise that this result does not invoke any global assumption on the spacetime and hence is valid for both asymptotically AdS and locally AdS spacetimes.  
\section{All three-dimensional solutions}
\label{sec:3d}

Three-dimensional Einstein-Maxwell theory with $\Lambda=-2 \ell^{-2}<0$ admits black hole solutions~\cite{Banados:1992wn, Clement:1992ke, Clement:1995zt, Martinez:1999qi}. It is easy to completely classify near-horizon geometries with compact cross-sections $S= S^1$~\cite{Kunduri:2013ana}. Being one-dimensional, $S$ has no curvature and all tensors are scalars, so we can introduce a periodic coordinate $x \sim x+2\pi R$ so that $\gamma^{(0)}=1$. Furthermore, the Maxwell field induced on $S$ must vanish so $B^{(0)}=0$.  There are two classes of near-horizon solutions: (i) AdS$_2\times S^1$; (ii) locally AdS$_3$~\cite{Kunduri:2013ana}.

For AdS$_2\times S^1$ the horizon data is $F^{(0)}= -2 \ell^{-2}, h^{(0)}=0, \Psi^{(0)}=\pm \ell^{-1}$. The linearised Einstein equation (\ref{einstein}) is automatically satisfied, whereas the Maxwell equation (\ref{maxwell}) reduces to 
\be
\frac{\td^2 Z^{(1)}}{\td x^2}- \tfrac{1}{2} \Psi^{(0)}\frac{\td  \gamma^{(1)}}{\td x}=0  \; .
\ee
Integrating we find the general solution
\be
Z^{(1)}(x)= a+ \tfrac{1}{2} \Psi^{(0)} \int^x \left(\gamma^{(1)}  - b \right)   \; ,
\ee
where $a,b$ are integration constants and periodicity of $Z^{(1)}$ fixes $b= \frac{1}{2\pi R} \int_{S^1} \gamma^{(1)}$. Thus the deformation is parameterised by a constant $a$ and an arbitrary function $\gamma^{(1)}$, reflecting the gauge freedom (\ref{gautransgamma}) which reduces to $\gamma^{(1)} \to \gamma^{(1)} + f''$.  The remaining  first order data is
\be
h^{(1)} = 2 \Psi^{(0)} Z^{(1)}, \qquad F^{(1)} = \tfrac{1}{3} (\Psi^{(0)})^2 b \; , \qquad   \Psi^{(1)} = -\tfrac{1}{2} \Psi^{(0)} b \; . 
\ee
The MTS condition is simply $b>0$. 
The static charged extreme BTZ solution corresponds to $a=0$ and $\gamma^{(1)}$ a  positive constant (so $b= \gamma^{(1)}>0$). The rotating generalisation~\cite{Clement:1992ke, Clement:1995zt, Martinez:1999qi} also has the AdS$_2\times S^1$ near-horizon geometry~\cite{Kunduri:2013ana} and presumably corresponds to the $a\neq 0, \gamma^{(1)}=b>0$ deformations.

The locally AdS$_3$ near-horizon geometry is given by the vacuum solution $F^{(0)}=0, h^{(0)} = \pm 2 /\ell, \Psi^{(0)}=0$. The linearised Einstein equation (\ref{einstein}) is again automatically satisfied, whereas the Maxwell equation (\ref{maxwell}) now reduces to 
\be
\frac{\td^2 Z^{(1)}}{\td x^2}- 3 h^{(0)} \frac{\td Z^{(1)}}{\td x} + 2 {h^{(0)2}} Z^{(1)}=0  \; .   \label{Z1eq3d}
\ee
Multiplying by $ \frac{\td Z^{(1)}}{\td x}$ the first and third terms become total derivatives, and integrating this over $S^1$ the boundary terms vanish (by periodicity), leaving
\be
\int_{S^1} \left( \frac{\td Z^{(1)}}{\td x}\right)^2 \td x =0  \; .
\ee
This implies $Z^{(1)}$ is a constant.  Substituting back into (\ref{Z1eq3d}) we deduce that $Z^{(1)}=0$. Therefore, the general deformation in this case is given by an arbitrary function $\gamma^{(1)}$  and the remaining first order data is
\be
h^{(1)} =  \tfrac{1}{4} h^{(0)} \gamma^{(1)} , \qquad F^{(1)}=0, \qquad \Psi^{(1)} = 0 \; .
\ee
In fact this is the general vacuum deformation\cite{Li:2015wsa}.  Therefore we find that at first order there are no electrovacuum deformations which are not vacuum. While charged black holes with an AdS$_2 \times S^1$ near-horizon geometry are known (as discussed above), we are not aware of any charged black holes with a locally AdS$_3$ near-horizon geometry (see discussion in~\cite{Kunduri:2013ana}). Indeed, such solutions may not exist and our result supports this possibility.

In the vacuum case, the full nonlinear solution to the Einstein equation is known for arbitrary $\gamma^{(1)}$ and is diffeomorphic to the extreme BTZ black hole (the diffeo is large, in the sense that the asymptotic Virasoro charges change)~\cite{Li:2013pra}. It would be interesting to find the full non linear solution in the Einstein-Maxwell case.  

\section*{Acknowledgements} JL would like to thank Gary Gibbons for useful comments. CL was supported by the Polish National Science Centre grant No. 2015/17/B/ST2/02871. JL is supported by STFC [ST/L000458/1].

\appendix

\section{Transverse deformations of  known solutions}

In this Appendix we compute the transverse deformations corresponding to the known extreme black holes solutions.  First, we present a general analysis for stationary and axisymmetric solutions. Then we apply it to various examples.

\subsection{Gaussian null coordinates for axisymmetric extreme black holes}
Consider a stationary and axisymmetric spacetime of the form
\bea
\label{genform}
\td s^2 = - R^2 A \td t^2+ \frac{B \td R^2}{R^2} + W\td x^2+ X (\td \phi + R \omega \td t)^2
\eea
where the metric components are functions of $(R, x)$. The surface $R=0$ is a smooth extremal horizon with normal $n=\partial_t$ if
\bea
B- (c- \tilde{c}R)^2 A= O(R^2) \; , \qquad \omega = \frac{b}{c} + O(R)  \; ,
\eea
where $b,c, \tilde{c}$ are constants (assume $c>0$).  These conditions can be written as
\bea
B_0(x)= c^2 A_0(x), \qquad B_1(x) = c^2 A_1(x) - 2 c \tilde{c} A_0(x), \qquad \omega_0(x) = \frac{b}{c}
\eea
where $B_n(x) = \partial^n_R B |_{R=0}$ etc. Indeed, in terms of the new coordinates $(V, \varphi)$ defined by
\be
t= V + \frac{c}{R} + \tilde{c} \log R, \qquad \phi = \varphi+ b \log R
\ee
we have
\bea
\td s^2 &=& A \left[- R^2  \td V^2+ 2(c- \tilde{c}R) \td V \td R \right]  + \frac{B- (c- \tilde{c}R)^2 A}{R^2} \td R^2 \nonumber \\ &+&  W\td x^2+ X \left( \td \varphi  + R \omega \td V + \frac{\td R}{R}( b- (c- \tilde{c}R) \omega) \right)^2
\eea
 and hence the above conditions imply $g_{R\mu} =O(1)$ near $R=0$ so that the spacetime metric is smooth and non-degenerate at $R=0$. In particular $R=0$ is an extremal Killing horizon.  Its near-horizon geometry can be extracted by scaling $(V, R) \to (V/ \epsilon, \epsilon R)$ and letting $\epsilon \to 0$, giving
 \be
 \td s_{\text{NH}}^2 = A_0[- R^2  \td V^2+ 2c\, \td V \td R]  +  W_0\td x^2+ X_0 \left( \td \varphi  + R \omega_0 \, \td V \right)^2  \; ,
\ee
which takes the familiar form of a circle fibration over AdS$_2$.
 
 We now find Gaussian null coordinates for extremal horizons of the above form.  The Killing fields are $n = \partial_t$ and $m= \partial_\phi$. We will assume the existence of an axisymmetric cross-section $S$, i.e.  such that $m$ is tangent to $S$. Then, we need to find null geodesics $\gamma(r)$ such that $\dot{\gamma} \cdot n=1$ and $\dot{ \gamma} \cdot m=0$, where $r$ is an affine parameter synchronised so that $r=0$ at the horizon. These give
 \be
 \dot{t}  = - \frac{1}{R^2 A}, \qquad \dot{\phi}= \frac{\omega}{R A}  \; .
 \ee
 The null condition is then
 \be
 - A^{-1} + B \dot{R^2} + R^2 W \dot{x}^2=0  \; ,
 \ee
 which together with the geodesic equation for $x$  and the initial conditions
 \be
 R(0,y)=0, \qquad x(0,y)=y, \qquad \dot{x}(0,y)=0
 \ee
uniquely  determines $R(r, y)$ and $x(r,y)$. We can develop the Taylor series in $r$ for the solution.  Using the null constraint we find
\be
R(r, y) = \frac{r}{c A_0(y)} \left(1   - \frac{(\tilde{c}A_0(y)- cA_1(y) )r}{2 c ^2A_0(y)^2} + O(r^2) \right) \; ,\qquad x(r,y) = y+ O(r^2) \; .  \label{gnc1}
\ee
Integrating for $t, \phi$ then gives
\bea
t &=& v+ \frac{c^2 A_0(y)}{r}+ \tilde{c} \log r+f(y)+ O(r), \nn \\  \phi &=& \psi+ g(y) + c \omega_0 \log r + \frac{r}{A_0(y)} \left( \omega_1(y)- \frac{\omega_0 A_1(y)}{2 A_0(y)} - \frac{\tilde{c} \omega_0}{2c} \right)+ O(r^2) \; ,   \label{gnc2}
\eea
where $v,\psi$ and $f(y),g(y)$ are constant along the geodesics.

The above defines a new chart $(v,r,y,\psi)$ near the horizon. The Killing fields in this chart are $n = \partial_v, m = \partial_\psi$. In order for this to define a Gaussian null chart we need to impose $g_{r y}=g_{r \psi}=0$. The latter condition is just $\partial_r \cdot m=0$ which we have already imposed. The former is
\be
\partial_r \cdot \partial_y =  - R^2 A \dot{t} t_y+\frac{B \dot{R} R_y}{R^2}+ \dot{x} x_y W = t_y+ \frac{B \dot{R} R_y}{R^2}  + O(r) \; ,
\ee
where $t_y = \partial_y t$ etc, 
which vanishes at $r=0$ if and only if
\be
f(y) = \frac{cA_1}{2A_0}  - \tilde{c} \log A_0  \; .
 \ee
 Since $\partial_r$ is geodesic this is sufficient to guarantee it vanishes for $r>0$. Hence, we have GNC.
 
 We have not fully fixed the coordinates on the horizon yet. Indeed,
 \be
 g_{y \psi} = X (\phi_y +R \omega t_y) \; ,
 \ee
 and requiring this to vanish on the horizon implies 
 \be
 g'(y) = - \frac{ c \omega_0 A_0'}{A_0}  \; .
 \ee
 Our coordinate change is now fully fixed. 
 
 We deduce the remaining components of the metric are 
 \bea
&&g_{yy} =\frac{R_y^2 B}{R^2} - R^2 A t_y^2+ x_y^2 W+ X( \phi_y+ R \omega t_y)^2 = W + O(r^2), \\  && g_{\psi\psi} = X, \qquad 
g_{vv} = - R^2 A + R^2 X \omega^2, \qquad g_{v\psi}=R X \omega  \\  && g_{yv} = - R^2 A t_y  + R X \omega (\phi_y + \omega R t_y)  \; .
\eea
From this we can extract the horizon data
\bea
\gamma^{(0)} = W_0 \td y^2 + X_0 \td \psi^2, \qquad h^{(0)} = \frac{\omega_0 X_0}{c A_0} \td \psi - \frac{A_0'}{A_0} \td y, \qquad F^{(0)} = \frac{ -A_0+ \omega_0^2 X_0}{ c^2 A_0^2}  \; .
\eea
The first order data $\gamma^{(1)}_{ab} = \partial_r \gamma_{ab}|_{r=0}$ is
\be
\gamma^{(1)}_{yy} = \frac{1}{c A_0(y)} W_1, \qquad \gamma^{(1)}_{\psi\psi} = \frac{1}{c A_0(y)} X_1, \qquad  \gamma^{(1)}_{y\psi} =  \frac{X_0 \omega_1'}{A_0}  \; .
\ee
Observe that for a static solution $\omega=0$ and hence $\gamma^{(1)}_{y \psi}=0$.

To apply the method to rotating black holes, we need to take account of the horizon rotating.  Thus we let $\tilde{t}=t$ and $\tilde{\phi} = \phi + \Omega_H t$, so
\bea
\td s^2 = - R^2 A \td \tilde{t}^2+ \frac{B \td R^2}{R^2} + W\td x^2+ X (\td \tilde{\phi} + \Omega \td \tilde{t})^2
\eea
where $\Omega \equiv  -\Omega_H + R \omega$.   Then the Killing field null on the horizon is $n = \partial_{\tilde{t}}+\Omega_H \partial_{\tilde{\phi}}$.

We now apply the above to work out the first order data for several important examples.

\subsection{Majumdar-Papapetrou solution} 
The Majumdar-Papapetrou solution is determined by an arbitrary harmonic function $H$ on $\mathbb{R}^3$. For a black hole at the origin of $\mathbb{R}^3$ the solution is
\be
\td s^2 = - H^{-2} \td t^2 + H^2 ( \td R^2 + R^2 \td \Omega_2^2) , \qquad H  = \frac{Q}{R} + \sum_{\ell=0}^\infty h_\ell R^\ell Y_\ell(\theta)
\ee
where we assume $H$ is axisymmetric (so the spacetime is).
Thus the metric takes the above general form (\ref{genform}) with
\be
A = \frac{1}{(Q + R \tilde{H})^2}, \qquad B = (Q+ R \tilde{H})^2,
\ee
where $\tilde{H}  = \sum_{\ell=0}^\infty h_\ell R^\ell Y_\ell(\theta)$. Hence $A_0 = 1/Q^2$ and $c= Q^2$, $\tilde{c} = -2 Q h_0$ and therefore
\be
\gamma^{(0)}= Q^2 \td \Omega_2^2, \qquad \gamma^{(1)} = 2 Q h_0 \td \Omega_2^2  \; .
\ee
Observe that the first order data in this case depends only the monopole term. Therefore to first order in GNC the multi-centred black hole solution is indistinguishable from a single black hole solution.

\subsection{Kerr-Newman-AdS} 
The Kerr-Newman-AdS solution, with $\Lambda= - 3g^2\leq 0$, in standard Boyer-Lindquist coordinates is given by 
\bea
\td s^2 &=& -\frac{\Delta_r}{\rho^2} \left( \td t - \frac{a (1-x^2)}{\Xi} \td \phi \right)^2 + \frac{\rho^2 \td r^2}{\Delta_r} \td r^2 + \frac{\rho^2}{(1-x^2) \Delta_x} \td x^2 \\ &&\quad + \frac{\Delta_x (1-x^2)}{\rho^2} \left( a \td t - \frac{r^2+a^2}{\Xi} \td \phi \right)^2  \; , \nn \\
\mcF &=& - \frac{1}{\rho^4} \left[ q_e (r^2 - a^2 x^2) + 2 q_m r a x \right] \left( \td t \wedge \td r + \frac{a (1-x^2)}{\Xi} \td r \wedge \td \phi \right) \\
\nn && \quad + \frac{1}{\rho^4} \left[ q_m (r^2 - a^2 x^2) - 2 q_e r a x \right] \left( a  \td t \wedge \td x + \frac{(r^2 +a^2)}{\Xi} \td x \wedge \td \phi \right)  \; ,
\eea
where 
\bea
\rho^2 &=& r^2 + a^2 x^2 \; , \quad \Xi= 1- g^2 a^2 \; , \quad \Delta_x = 1 - g^2a^2 x^2 \\ \Delta_r &=& (r^2 + a^2)\left(1 + g^2 r^2 \right) - 2mr + z^2 \; .
\eea
The parameters $m,a$ encode the mass and rotation, whereas $q_e$ and $q_m$ are electric and magnetic charges respectively and $z^2 = q_e^2 + q_m^2$.
The horizon is located at the biggest root $r_+>0$ of $\Delta_r =0$. In the extreme limit the parameters obey
\bea
m &=& \frac{g^2 r_+^4 + a^2 (1 + g^2 r_+^2) + (r_+^2+ z^2)}{2 r_+}, \\ z^2 &=& 3 g^2 r_+^4 + \left(1 +  g^2 a^2 \right)  r_+^2 - a^2 
\eea 
and 
\bea
\Delta_r = (r-r_+)^2[1+ g^2(r^2 + 2 r r_+ + a^2 + 3 r_+^2)]  \; .
\eea
The Kerr-Newman case in given by setting $g=0$ in the above, in which case $r_+^2=a^2+z^2$. The Kerr-AdS case is given by setting $z=0$, in which case  $a^2 = (1+3 g^2 r_+^2)/(1-g^2 r_+^2)$ and $g^2r_+^2<1$.

Writing the metric in our general form (\ref{genform}) and setting $R= r-r_+$ we can read off the horizon data.
One finds the constants are
\bea
\Omega_H &=& \frac{a \Xi}{r_+^2 +a^2}, \qquad \omega_0=  \frac{2 ar_+ \Xi}{(r_+^2+a^2)^2} \; ,\\
 c &=& \frac{r_+^2+a^2}{1+g^2 a^2+ 6 g^2 r_+^2}, \qquad \tilde{c}= -\frac{2r_+ \left(-a^2 g^2+4 g^2 r_+^2+1\right)}{\left(a^2 g^2+6 g^2r_+^2+1\right)^2} \; ,
\eea
and the horizon data is
\bea
A_0 &=& \frac{\left(a^2 g^2+6 g^2 r_+^2+1\right)\rho_+^2 }{\left(a^2+r_+^2\right)^2} \\
\gamma^{(0)} &=& \frac{\rho_+^2 \td x^2}{\left(1-x^2\right)\Delta_x }  +  \frac{\left(1-x^2\right) \left(a^2+r_+^2\right)^2\Delta_x \td \phi^2}{\Xi^2\rho_+^2}\    \; ,
\eea
where $\rho_+^2= r_+^2+ a^2 x^2$. 
The first order data is
\bea
\gamma^{(1)}_{xx} &=& \frac{2 r_+ \left(a^2+r_+^2\right)}{\left(1-x^2\right)\Delta_x\rho_+^2} \\
\gamma^{(1)}_{x\phi} &=& \frac{2 a^3 x \left(1-x^2\right) \left(a^2+r_+^2\right) \left(g^2 r_+^2+1\right)}{\Xi \Delta_x \rho_+^4} \\
\gamma^{(1)}_{\phi\phi} &=& \frac{2 r_+ \left(1-x^2\right) \left(a^2+r_+^2\right)^2\Delta_x \left(a^2 \left(2 x^2-1\right)+r_+^2\right)}{\Xi^2 \rho_+^6}  \; .
\eea
The near-horizon Maxwell field and deformation can be computed using the coordinate change to GNC (\ref{gnc1}), (\ref{gnc2}).  We find
\bea
\Psi^{(0)} &=& \frac{a^2q_e x^2-2 aq_m r_+ x-q_e r_+^2}{\rho_+^4}\\
{B}^{(0)}_{x\phi} &=&- \frac{\left(a^2+r_+^2\right) \left(a^2 q_m x^2+2 a q_e r_+ x-q_m r_+^2\right)}{\Xi \rho_+^4}
\eea
and
\bea
Z^{(1)}_x &=& -  \frac{a [ q_m (r^2_+-a^2 x^2) -2 a q_e r_+ x]}{\Xi\rho_+^4} \\
Z^{(1)}_\phi &=& \frac{a \left(1-x^2\right) \left(a^2+r_+^2\right) \left(a^2 q_e x^2-2 a q_m r_+ x-q_e r_+^2\right)}{\Xi \rho_+^6}   \; .
\eea

\subsection{Kerr-Newman-Melvin}

The Kerr-Newman-Melvin spacetime may be constructed by applying a Harrison transformation to the Kerr-Newman metric, see eg.~\cite{Gibbons:2013yq}. In particular,  the extreme limit has been studied in detail in recent years~\cite{Booth:2015nwa, Hejda:2015gna, Bicak:2015lxa}. The extreme solution is a three parameter family which depends on $(\tilde{a}, \tilde{q}, b )$, where $\tilde{a}, \tilde{q}$ are the rotation and charge parameter of the seed Kerr-Newman solution and $b$ parameterises the external magnetic field.\footnote{If one includes magnetic charge in the seed Kerr-Newman, one gets conical singularities on the poles of horizon. Removing these again gives a three parameter family of solutions. We will not consider this solution.}

The extreme solution can be written as
\bea
\td s^2= |\Lambda |^2 \rho^2 \left( - \frac{\Delta}{\mathcal{A}} \td t^2+ \frac{\td r^2}{\Delta}+ \frac{\td x^2}{1-x^2} \right) + \frac{\mathcal{A}  | \Lambda |^2 (1-x^2) }{\rho^2}( \td \phi- \hat{\omega} \td t)^2
\eea
where
\be
\Delta = (r-m)^2, \qquad \rho^2=r^2+\tilde{a}^2 x^2, \qquad \mathcal{A} = (r^2+\tilde{a}^2)^2 - \Delta \tilde{a}^2 (1-x^2) \;  ,
\ee
the parameter $m= \sqrt{\tilde{a}^2+ \tilde{q}^2}>0$ and $\Lambda, \hat{\omega}$ are functions that depend on the magnetic field parameter $b$. The general solution is rather complicated. For simplicity we will consider two special cases. The Reissner-Nordstr\"om-Melvin solution is given by $\tilde{a}=0$ and arises as a special case of the rotating Ernst solution we give in the next section.

The extreme Kerr-Melvin solution is given by $m=\tilde{a}$ and the functions
\bea
\Lambda &=& 1+ \tfrac{1}{4}  \frac{b^2\mathcal{A} (1-x^2) }{\rho^2} - \tfrac{1}{2} i \tilde{a}^2 x \left( 3- x^2 + \frac{\tilde{a}^2}{\rho^2}(1-x^2)^2 \right) \\
\hat{\omega}&=& \frac{\tilde{a}}{r^2+\tilde{a}^2} \left\{ (1- b^4 \tilde{a}^4) - \Delta \left[ \frac{\rho^2}{\mathcal{A}} + \frac{b^4}{16} \left( - 8 \tilde{a} r x^2(3-x^2)- 6 \tilde{a} r (1-x^2)^2 \right. \right.  \right.  \\  && \left. \left. \left. + \frac{2 \tilde{a}^3 (1-x^2)^3}{\mathcal{A}}[ r(r^2+\tilde{a}^2)+2 \tilde{a}^3] + \frac{4 \tilde{a}^4 x^2}{\mathcal{A}} \left[ (r^2+\tilde{a}^2) (3-x^2)^2 - 4 \tilde{a}^2 (1-x^2) \right] \right) \right] \right\} \nn
\eea
where $b$ is the magnetic field parameter.
Setting $R=r-\tilde{a}$ we may extract the constants
\bea
\Omega_H = \frac{1- \tilde{a}^4 b^4}{2\tilde{a}}, \qquad \omega_0 = \frac{1- \tilde{a}^4 b^4}{2\tilde{a}^2}, \qquad
c= 2 \tilde{a}^2, \qquad \tilde{c}=-2 \tilde{a}
\eea
and the horizon geometry is that of extreme Kerr-Newman (see previous section with $g=0$) with
\bea
r_+^2 &=& \tilde{a}^2(1+ \tilde{a}^2 b^2)^2, \qquad a^2= \tilde{a}^2(1- \tilde{a}^2 b^2)^2 \\ 
A_0 &=& \frac{r_+^2+ a^2 x^2}{4 \tilde{a}^4}  \; .
\eea
The first order data is
\bea
\gamma^{(1)}_{xx} &=& \frac{4 \tilde{a}^3 \left[ (1+\tilde{a}^2 b^2)^2-2 \tilde{a}^2 b^2 x^2\right]}{\left(1-x^2\right) (r_+^2+a^2 x^2)} \\
\gamma^{(1)}_{\phi \phi} &=& \frac{16 \tilde{a}^7 x^2 \left(1-x^2\right) \left(1+\tilde{a}^4 b^4 \right)}{\left(r_+^2+a^2 x^2\right)^3} \\
\gamma^{(1)}_{x\phi} &=& \frac{4 \tilde{a}^5 x \left(1-x^2\right) \left(1- \tilde{a}^4 b^4\right)}{\left(r_+^2+ a^2 x^2\right)^2}  \; .
\eea
Observe that for $b=0$ this reduces to the extreme Kerr data. For $\tilde{a}^2 b^2=1$ the near-horizon geometry is static so this gives a deformation of Ad$S_2\times S^2$; in fact it is a static deformation, although the full solution is not static.

\subsection{Ernst solution and rotating generalisation} 
\subsubsection{Ernst solution}
The Ernst solution represents a static charged accelerating black hole held in equilibrium by an external field, see e.g.~\cite{Griffiths:2009dfa}. This is given by
\bea
\td s^2= \frac{1}{(1+A r \tilde{x})^2} \left[ D^2 \left( - Q \td t^2+ \frac{\td r^2}{Q}+ \frac{r^2 \td \tilde{x}^2}{P (1-\tilde{x}^2)} \right)+ \frac{P r^2(1-\tilde{x}^2) \td \tilde{\phi}^2}{D^2} \right]
\eea
where in the extreme limit
\bea
&&Q= r^{-2}(r-e)^2 (1-A^2 r^2), \qquad P = (1+A e \tilde{x})^2 \; ,\\  &&D = (1+ \tfrac{1}{2} e b \tilde{x})^2+ \frac{b^2 r^2 (1-\tilde{x}^2) P}{4 (1+A r \tilde{x})^2}  \; .
\eea
The parameter $e$ represents the electric (or magnetic) charge, $b$ the external electric (or magnetic) field and $A$ the acceleration. The black hole horizon is at $r=e$ and there is an acceleration horizon at $r=1/|A|$. We require the acceleration horizon to be outside the black hole so $e^2 A^2<1$ and $e<r<1/|A|$.
The conical singularities at $\tilde{x}= \pm 1$ can be simultaneously removed if
\be
A=  \frac{b}{1+ \tfrac{1}{4} e^2 b^2}
\ee
and the period of $\tilde{\phi}$ is chosen appropriately (see below). The condition $e^2 A^2<1$ thus implies $e^2b^2<4$.  For $b=0$ this solution reduces to the extreme Reissner-Nordstr\"om solution.

Setting $R= r-e$ we may use the above general formulas to extract the horizon data and the first order deformation.  We find the horizon data
\bea
&&A_0 = \frac{(4- b^2 e^2)^2}{16 e^2}, \qquad c = \frac{ e^2(4+ b^2 e^2)^2}{(4-b^2 e^2)^2},  \qquad \tilde{c} = - \frac{2 e (4+ b^2 e^2)^4}{(4- b^2 e^2)^4} \\
&&\gamma^{(0)} =r_+^2\left( \frac{ \td x^2}{1-x^2} + (1-x^2)  \td \phi^2 \right)  \; ,
\eea
where 
\be
r_+= \frac{e (4+ b^2 e^2)^2}{ 4( 4- b^2 e^2)} \; ,
\ee
 and to reveal the round metric on the horizon we have changed coordinates to ($x, \phi$)
\be
\tilde{x}= \frac{x- A e}{1- A e x}, \qquad  \tilde{\phi}= Z^{-1} \phi  , \qquad Z = \frac{16}{(4+ b^2 e^2)^2} \; ,
\ee
where $\phi$ is $2\pi$ periodic. 
We then find that the first order data is
\bea
&&\gamma^{(1)}_{{\phi}{\phi}} = \frac{512 e( 1-x^2)[4-4 be x+b^2 e^2(-1+2 x^2)]}{(4-b^2 e^2)^4 (4+b^2 e^2)}  \\
&&\gamma^{(1)}_{xx}= \frac{2 e(4+ b^2 e^2)^3[ 4-4 be x- b^2 e^2(-3+2 x^2)]}{(4-b^2 e^2)^4(1-x^2)} \\
&&\gamma^{(1)}_{x\phi}=0  \; .
\eea
Notice that $\gamma^{(1)}_{x\phi}=0$ is  due to the fact the solution is static. For $b=0$ this reduces to $\gamma^{(1)}= 2 e \, \td \Omega_2^2$, corresponding to the Reissner-Nordstr\"om solution.
\subsubsection{Rotating generalisation}

A magnetised accelerating Kerr-Newman solution can be obtained by magnetising the accelerating Kerr-Newman via the Ehlers-Harrison transformation. The seed extremal accelerating Kerr-Newman metric can be written as~\cite{Griffiths:2009dfa}
\bea
\nn \td s^2 &=&  \frac{1}{\left( 1+ r \tilde{x} A \right)^2}  \left\lbrace -\frac{G(r)}{\rho^2} \left[\td t + \tilde{a} \left(1-\tilde{x}^2\right)\td \tilde{\phi}\right]^2 + \frac{\rho^2}{G(r)} \td r^2 \right.  \\
&+& \left. \frac{H(\tilde{x})}{\rho^2} \left[\left(r^2 + \tilde{a}^2 \right)  \td \tilde{\phi} + a \td t \right]^2 + \frac{\rho^2}{H(\tilde{x})} \td \tilde{x}^2 \right\rbrace \label{AKN} \; , 
\eea
where 
\beas
&&  \rho^2 = r^2 + \tilde{a}^2 \tilde{x}^2 \; ,\\ 
&& G(r):= \left(1-A^2 r^2\right) \left( r- m \right)^2 \; , \\
&& H(\tilde{x}) := \left(1-\tilde{x}^2 \right) \left( 1+ A \tilde{x} m \right)^2 \; ,
\eeas
and $m=\sqrt{\tilde{a}^2 + \tilde{q}^2 + \tilde{p}^2 }>0$. 
The black hole horizon is located at $r = m$ and there is also an acceleration horizon at $r= r_A= \frac{1}{|A|}$. Thus for the acceleration horizon to be outside the black hole horizon, we must have $r_A > m$ and so $A^2 m^2<1 $. The coordinate ranges are $m<r<r_A$, $-1< \tilde{x}<1$. For $A \neq 0$ this metric has conical singularities at $\tilde{x}=\pm 1$ which cannot be simultaneously removed.  For $A=0$ the solution reduces to the extremal Kerr-Newman black hole.

For simplicity we will set $\tilde{a}=0$; for $\tilde{q}\neq 0$ this still leads to a rotating solution due to the presence of the external magnetic field. The magnetised solution is given in~\cite{Astorino:2016xiy} and takes the form
\bea
\nn \widehat{\td s^2} &=&  \frac{| \Lambda |^2}{\left( 1+ r \tilde{x} A \right)^2}  \left\lbrace -\frac{G(r)}{r^2} \td t^2  + \frac{r^2}{G(r)} \td r^2 +  \frac{r^2}{H(\tilde{x})} \td \tilde{x}^2 \right\rbrace  \\ &+&  \frac{r^2 H(\tilde{x})}{| \Lambda |^2 \left( 1+ r \tilde{x} A \right)^2} \left(\td \tilde{\phi} -\hat{\omega} \td t \right)^2 \label{AKN} 
\eea
where
\bea
\Lambda &=& 1+ b \tilde{x} (\tilde{p}-i \tilde{q})+ \tfrac{1}{4} b^2 \left[ \frac{r^2H(\tilde{x})}{(1+A r \tilde{x})^2} + (\tilde{p}^2+\tilde{q}^2) \tilde{x}^2 \right] \\
\hat{\omega} &=& -\frac{2\tilde{q} b}{r} + \frac{\tilde{q} b^3\left[ (r^2-2 m r)(1+A r \tilde{x}+\tilde{x}^2) + \tilde{x}^2 (\tilde{p}^2+\tilde{q}^2)(1-A^2 r^2) \right]}{2 r (1+A r \tilde{x})^2}
\eea
and the parameter $b$ encodes the background magnetic field (so $b=0$ reduces to the above seed). We find it convenient to introduce the parameterisation $\tilde{q}= z \sin \alpha$ and $\tilde{p}= z\cos \alpha$, so that $m=z$ and hence $A^2 z^2<1$. In general the solution has conical singularities on the axis of symmetry $\tilde{x} = \pm 1$.  For $b \neq 0$, simultaneous removal of these can be achieved by setting\footnote{We find that this is simpler than solving for $A$ as was done in~\cite{Astorino:2016xiy}.}
\be
\cos \alpha = \frac{A (16+24 b^2 z^2+b^4 z^4)}{4 b (1+A^2 z^2)(4+b^2 z^2)}
\ee
and fixing the period to be
\be
\Delta \tilde{\phi} = \frac{2\pi}{Z}, \qquad Z = \frac{16(1+A^2 z^2)}{16+24 b^2 z^2 + b^4 z^4} \; .
\ee
This gives a 3-parameter family of solutions parameterised by $(z,A,b)$.   

The parameter ranges can be obtained as follows. From the expression for $\cos \alpha$, we have
\be
\sin^2\alpha = \frac{\left[ (4+b^2z^2)^2- 16 b^2 A^2 z^4 \right] \left( \frac{16 b^2}{(4+b^2z^2)^2} - A^2 \right)}{16 b^2(1+A^2 z^2)^2}  \; .
\ee
 The condition $A^2z^2<1$ mentioned above implies
\be
(4+b^2 z^2)^2 - 16 b^2 A^2 z^4>(4+b^2 z^2)^2- 16 b^2 z^2=(4-b^2 z^2)^2\geq 0 \; , 
\ee
and hence $\sin^2\alpha \geq 0$ implies
\be
A^2\leq \frac{16 b^2}{(4+b^2 z^2)^2}  \; . 
\ee
For $A=0$ the solution reduces to Reissner-Norstrom-Melvin with parameters $\tilde{q},b$ (note $\cos\alpha=0$ in this case so $z= \pm \tilde{q}$). For $A= \pm 4 b/(4+b^2z^2)$ we have $\tilde{q}=0$ (note $\cos\alpha= \pm 1$), which corresponds to the static Ernst solution.

Setting $R=r-m$ we may write this solution in our general form (\ref{genform}) to extract the near-horizon data.   We find the constants
\bea
\Omega_H &=& \frac{b q (4+ b^2 z^2)}{2 z},  \qquad \omega_0 = \tfrac{1}{2} b q \left( b^2+ \frac{4}{z^2} \right) \\
c  &=& \frac{z}{1-A^2 z^2}, \qquad \tilde{c} =- \frac{1+A^2 z^2}{(1-A^2 z^2)^2}  \; ,
\eea
thus establishing regularity of the event horizon.  The horizon data can be written as
\bea
\gamma^{(0)} &=& \frac{r_+^2+ a^2 x^2}{1-x^2} \td x^2 + \frac{ (1-x^2)(r_+^2+a^2)^2}{(r_+^2+ a^2 x^2)^2} \td \phi^2 \\
A_0 &=& \frac{(1-A^2 z^2)^2}{z^2} (r_+^2+ a^2 x^2)  \; ,
\eea
where we have defined the constants
\be
r_+^2 = \frac{ z^2 [ (4+b^2 z^2)^2- 16 A^2 b^2 z^4]}{16 (1- A^2 z^2)^2 (1+A^2 z^2)}, \qquad a^2 = \frac{ z^4 [ 16b^2- A^2(4+b^2 z^2)^2]}{16 (1- A^2 z^2)^2 (1+A^2 z^2)} \; ,   \label{rotErnstparam}
\ee
and changed coordinates to
\be
x = \frac{\tilde{x}+ A z}{1+ A z \tilde{x}}, \qquad  \phi =Z  \tilde{\phi}  \; .
\ee
This reveals the horizon geometry is isometric to that of the extreme Kerr-Newman (as it must be!), with parameters $r_+,a$ as given above.  Note that the parameter ranges ensure that $r_+^2>0$ and $a^2 \geq 0$. Note that a static near-horizon geometry $a=0$ corresponds to the static Ernst solution.

The first order deformation takes our general form with
\bea
A_1&=& \frac{A r_+ z^7 (4- b^2 z^2) \left(b^4 z^4+24 b^2 z^2+16\right) (A^2 \left(b^2 z^2+4\right)^2-16 b^2 )}{256 a (1- A^2 z^2)^5 \left(A^2 z^2+1\right)^2 \left(b^2 z^2+4\right)}   \label{rotErnstAi} \\
A_2 &=& \frac{z^4 (4- b^2 z^2) \left(b^4 z^4+24 b^2 z^2+16\right) \left(b^2 \left(8 z^2-16 A^2 z^4\right)+b^4 z^4+16\right)}{256 r_+ (1- A^2 z^2)^5 \left(A^2 z^2+1\right)^2 \left(b^2 z^2+4\right)}   \nn \\
A_3 &=& \frac{3 z^8 \left(b^4 z^4+24 b^2 z^2+16\right)^2 \left(- A^2 \left(b^6 z^6+44 b^4 z^4+48 b^2 z^2+64\right)+2 b^2 \left(b^4 z^4+16 b^2 z^2+48\right)\right)}{2048 (1-A^2 z^2)^6 \left(A^2 z^2+1\right)^3 \left(b^2 z^2+4\right)} \nn
\eea
The MTS condition
\be
A_3+ 6 A_2 r_+ (r_+^2+a^2) = \frac{3 z^6 (16+ 24 b^2 z^2+ b^4 z^4)^3}{2048(1-A^2 z^2)^5 (1+A^2 z^2)^3}>0
\ee
is satisfied.  For $A=0$ the above gives the first order data for the Reissner-Nordstr\"om-Melvin solution given in the main text in equation (\ref{ARNM}).

\end{document}